\documentclass[fleqn,usenatbib]{mnras}

\usepackage{newtxtext,newtxmath}

\usepackage[T1]{fontenc}

\DeclareRobustCommand{\VAN}[3]{#2}
\let\VANthebibliography\thebibliography
\def\thebibliography{\DeclareRobustCommand{\VAN}[3]{##3}\VANthebibliography}

\usepackage{graphicx}	
\usepackage{amsmath}	

\usepackage{amssymb}	

\usepackage{threeparttable, tablefootnote}

\title[X-ray emission in RLQ J1214+1402]{X-ray emission of radio-loud quasar SDSS J121426.52+140258.9: independent variations between optical/UV and X-ray emission}

\author[M. Zhou et al.]{
Minhua Zhou,$^{1}$\thanks{E-mail: zhoumh8@163.com}
Minfeng Gu,$^{2}$\thanks{E-mail: gumf@shao.ac.cn}
Mai Liao,$^{3,4}$
Muhammad S. Anjum,$^{5}$
\\
$^{1}$School of Physics and Electronic Information, Shangrao Normal University, 401 Zhimin Road, Shangrao 334001, China\\
$^{2}$Key Laboratory for Research in Galaxies and Cosmology, 
	Shanghai Astronomical Observatory, 
	Chinese Academy of Sciences, 80 Nandan Road, \\
	Shanghai 200030, China\\
$^{3}$CAS Key Laboratory for Research in Galaxies and Cosmology, 
Department of Astronomy, University of Science and Technology of China,\\
  Hefei, Anhui 230026, China\\
$^{4}$School of Astronomy and Space Science, University of Science and Technology of China, Hefei 230026, China\\
$^{5}$Department of Physics, Xiamen University Malaysia, Jalan Sunsuria, Bandar Sunsuria 43900 Sepang, Malaysia
}

\date{Accepted XXX. Received YYY; in original form ZZZ}

\pubyear{2021}

\begin{document}
\label{firstpage}
\pagerange{\pageref{firstpage}--\pageref{lastpage}}
\maketitle

\begin{abstract}

To understand the X-ray emission of active galactic nuclei (AGNs), we explored the optical-to-X-ray variation correlation of a radio-loud quasar (RLQ) SDSS J121426.52+140258.9 (hereafter J1214+1402) with multi-epoch observations of Swift and XMM-Newton telescopes.
With the historical multi-band data, we found that the infrared to X-ray flux of RLQ J1214+1402 should not be dominated by the beamed jet emission. 
The Swift optical/UV and X-ray light curves showed that J1214+1402 has two optical states with low flux before 2014 April 08 and high flux after 2014 June 11, but has no significant X-ray variations during the time range between 2007 March 09 and 2014 August 04. This result was supported by the XMM-Newton observations in the overlapped time with Swift.
Interestingly, the early XMM-Newton data prior to the Swift time presents two unusual emission epochs when J1214+1402 has relatively low optical fluxes but has the brightest X-ray fluxes. The overall independence of optical-to-X-ray variation seems hard to be described by the disk-corona model. With the X-ray spectral fitting, we find that the soft X-ray excess in J1214+1402 appears only during the high optical state when the X-ray emission is at low state.
The soft X-ray excess in J1214+1402 is difficult to be explained by the ionized accretion disk, instead, it may be related to the warm corona.

\end{abstract}

\begin{keywords}
quasars: individual (SDSS J121426.52+140258.9) -- X-rays: general -- galaxies: active -- radiation mechanisms: general
\end{keywords}



\section{Introduction}
\label{sec:intro}

With the radio and optical observations, active galactic nuclei (AGNs) can be classified into radio-loud and radio-quiet AGNs based on radio loudness \cite[the ratio of radio to optical luminosity, $R=f_{5\rm GHz}/f_{4400}$,][]{1989AJ.....98.1195K}, with $R \ge 10$ and $R<10$ in the former and latter, respectively. These two types of AGNs have similar infrared to optical/ultraviolet (UV) spectral energy distributions (SEDs), but show significant differences in radio and X-ray emission \cite[e.g.,][]{1994ApJS...95....1E, 2011ApJS..196....2S}. 
Radio-loud AGNs (RL-AGNs) statistically show higher radio luminosity than radio-quiet AGNs (RQ-AGNs), that is most likely due to the presence of powerful relativistic jets \cite[][]{1995PASP..107..803U}. 
However, the reason for the emission difference in the X-ray band between RL-AGNs and RQ-AGNs is still under debate \cite[e.g.,][]{1987ApJ...313..596W, 2002MNRAS.332L..45B, 2011ApJ...726...20M, 2018MNRAS.480.2861G, 2020MNRAS.492..315G, 2020MNRAS.497..482L, 2020ApJ...893...39Z, 2020MNRAS.496..245Z, 2021MNRAS.505.1954Z, 2021RAA....21....4Z, 2021MNRAS.503.1987C, 2021RNAAS...5..101T}. 

It is now widely accepted that the X-ray emission in RQ-AGNs originates in the immediate vicinity of the supermassive black hole, and contains several components including a primary X-ray emission with a high energy cutoff, soft X-ray excess, reflection, and absorption components \cite[e.g.,][]{2009A&ARv..17...47T, 2015A&ARv..23....1B}. The primary X-ray continuum emission arises via Compton up-scattering in an accretion disk ``corona'', and may then interact (or be obscured) with matter to produce reflected/scattered X-ray emission via Compton ``reflection'' and scattering \cite[e.g.,][]{2009A&ARv..17...47T, 2018ARA&A..56..625H}. Previous sample analysis showed that RL-AGNs are more X-ray luminous than RQ-AGNs \cite[e.g.,][]{2011ApJ...726...20M, 2018MNRAS.480.2861G, 2021RAA....21....4Z}. 
Investigating the higher X-ray flux in RL-AGNs,  \cite{1987ApJ...313..596W}, and \cite{2011ApJ...726...20M} argued that RL-AGNs may have an additional jet component at UV and X-rays, however, \cite{2002MNRAS.332L..45B} proposed that the accretion disk in RL-AGNs could be more ionized than that in RQ-AGNs. \cite{2018MNRAS.480.2861G, 2020MNRAS.492..315G} found that RL- and RQ-AGNs have similar X-ray spectral indices, and the Type 1 and Type 2 RL-AGNs have similar distribution of X-ray loudness. The authors argued that RL-AGNs and RQ-AGNs may have the same X-ray emission mechanism and the larger X-ray luminosity in RL-AGN may result from larger radiative efficiencies of the innermost portions of the accretion flows. 
This result of the similar coronal X-ray emission in RL- and RQ-AGNs was further supported by several recent studies, e.g., \cite{2020MNRAS.496..245Z, 2021MNRAS.505.1954Z, 2021MNRAS.503.1987C, 2021RNAAS...5..101T}. 

On the other hand, there are a series of works suggesting that the RL-AGNs may have jet-associated X-ray flux. \cite{2020MNRAS.497..482L} investigated the X-ray emission of a sample of young radio AGNs, and found the good dependence of radio on X-ray luminosity with coefficient equals about 1, which deviates from the theoretical prediction of accretion flow as the origin of X-ray emission. In contrast to RQ-AGNs \cite[e.g.,][]{2013MNRAS.433..648F, 2015A&ARv..23....1B}, the X-ray photon index of RL-AGNs does not correlate with the Eddington ratio \cite[][]{2019MNRAS.490.3793L, 2020ApJ...893...39Z, 2020MNRAS.497..482L, 2021MNRAS.505.1954Z, 2021RAA....21....4Z}. By dividing the radio-detected quasars into radio loudness bins, \cite{2021RAA....21....4Z} found a unified multi-correlation between the radio luminosity, X-ray luminosity, and radio loudness. 
The Chandra X-ray imaging shows that the jet knot, hotspot, and lobe may all contribute to X-ray emission \cite[][]{2006ARA&A..44..463H, 2015ApJS..220....5M}, see more examples in XJET\footnote{\url{http://hea-www.harvard.edu/XJET/}} project \cite[][]{2010AIPC.1248..355H, 2010AIPC.1248..475M, 2011ApJS..197...24M, 2011IAUS..275..160M}. 
\cite{2017ApJ...835..226K} found that the radio Eddington luminosity inversely scales with the X-ray reflection fraction and positively scales with the distance between the corona and the reflected regions in the disk, and explained these results with a moving corona-jet model. 
All these results imply different origins of X-ray emission in RL-AGNs and RQ-AGNs, and the X-ray emission of some RL-AGNs could be explained with the corona-jet model. 
If the X-ray emission in RL-AGNs is affected by (or originates from) the jet, the multi-band flux variations may be different for RQ-AGNs and RL-AGNs, such as the optical to X-ray correlation and/or time delay between them.

Variability is one of the general properties of AGNs \cite[see, e.g.,][]{1997iagn.book.....P, 2012agn..book.....B}, and it is widely used to understand the central physics of AGNs. 
The previous works on radio-quiet quasars (RQQs) and blazars found the main results about the AGN variability \cite[e.g.,][]{1997ARA&A..35..445U}, such as the flux correlation between optical/UV and other wavelengths \cite[ e.g.,][]{1991ApJ...371..541K, 1998ApJ...505..594N, 2000ApJ...535...58K, 2001ApJ...561..146C} and the wavelength-dependent variability, such that the AGNs show larger amplitude of variability for high energy bands as compared to the low energy ones \citep[see, e.g.,][]{1997A&A...321..123C, 2004ApJ...601..692V, 2012ApJ...744..147S}. 

In blazars, generally various multi-band correlations have been found \cite[e.g., radio-optical-X-ray flux correlation, ][]{2008A&A...486..411S, 2008ApJ...689...79C, 2015A&A...576A.126A}. These correlations support the model in which the low-frequency and high-frequency continuum emissions arise from the same population of relativistic electrons by synchrotron or inverse Compton processes \cite[][]{1997ARA&A..35..445U}. 
The correlation of variability between optical/UV and X-ray bands has also been found in RQQs \cite[e.g.,][]{2009MNRAS.394..427B, 2010MNRAS.403..605B, 2012MNRAS.422..902C, 2017MNRAS.466.1777P, 2017ApJ...840...41E} and it can be well described by the reprocessing model in which the accretion disk is illuminated by the variable central X-ray flux \cite[][]{1988MNRAS.233..475G, 1997ARA&A..35..445U}. However, \cite{2018ApJ...860...29Z} and \cite{2018ApJ...868...58K} have disfavored the disk reprocessing scenario in the context of variability correlation. \cite{2018ApJ...868...58K} associated the variation of optical/UV emission with a magnetic turbulence model.  

However, to our knowledge, there are only a few relevant works on the variability of radio-loud quasars (not including blazars as in our work). 
\cite{2009ApJ...704.1689C, 2011ApJ...734...43C} studied the variability of RL-AGNs 3C 120 and 3C 111 with multi-band data and found a good correlation between optical/UV and X-ray bands. However, \cite{2009A&A...495..691M} and \cite{2014MNRAS.439..690H} classified 3C 120 and 3C 111 as blazars, respectively. No flux correlation between X-rays and the R band has been found for the broad-line radio quasar 4C +74.26 \cite[][]{2018ApJ...866..132B}. 
More RLQs need to be studied for a multi-band correlation. The study of an individual RLQ by analyzing its simultaneous optical to X-ray data may reveal the intrinsic X-ray emission mechanism in quasars. 

The XMM-Newton and the Neil Gehrels Swift Observatory (hereafter Swift) provide the ideal simultaneous optical to X-ray data. In this work, we chose RLQ SDSS J121426.52+140258.9 (hereafter J1214+1402) to study the multi-epoch simultaneous optical/UV to X-ray observations by Swift and XMM-Newton telescopes, see Section \ref{subsec:swift} and \ref{subsec:newton} for details on Swift and XMM-Newton data, to investigate the emission mechanism of X-rays in RL-AGNs.  
In the following Section, the results of multi-band and X-ray data of J1214+1402 are presented. 
Discussion and conclusion are organized in Section \ref{sec:discus} and \ref{sec:sum}, respectively.
Throughout this paper, the spectral index $\alpha$ is defined by $f_{\rm \nu} \varpropto \nu^{-\alpha}$, where $f_{\rm \nu}$ is the flux density at frequency $\nu$, and photon index $\Gamma$ is defined by $A(E)=KE^{-\Gamma}$, where $K$ is the power-law normalization at $1~\rm keV$ and $E$ is X-ray photon energy. We assume the flat cosmology model with $H_{0}=70\rm ~km~ s^{-1}~Mpc^{-1}$, $\Omega_{\rm m}=0.3$, $\Omega_{\Lambda}=0.7$, and the luminosity distance of J1214+1402 is about 9089 $\rm Mpc$.

\section{Data reduction and Results}
\label{sec:J1214}

J1214+1402 ($\alpha_{2000}=\rm 12^h14^m26.525^s$, $\delta_{2000}=\rm +14^d02^m58.91^s$, $z=1.280$) is a RLQ with radio loudness $ R=4029.1 $ \cite[][]{2021RAA....21....4Z}, calculated using the data from FIRST \cite[][]{1995ApJ...450..559B} and Sloan Digital Sky Survey \cite[SDSS, ][]{2017AJ....154...28B} catalogs. 

\subsection{Multi-band Data}\label{subsec:Mul-info}

In the optical band, J1214+1402 was spectroscopically observed by the SDSS \cite[][]{2000AJ....120.1579Y} on 2015 April 7. By SDSS optical spectral analysis \cite[][]{2011ApJS..194...45S}, J1214+1402 contains a supper massive black hole with black hole mass $\log M_{\rm BH}=9.23 \pm 0.09~\rm [M_\odot]$, bolometric luminosity $\log L_{\rm bol}=46.24\pm 0.01~\rm [erg/s]$, and the corresponding Eddington ratio $R_{\rm Edd}=0.081$.

\begin{figure*}
	\centering
	\includegraphics[width=0.95\columnwidth]{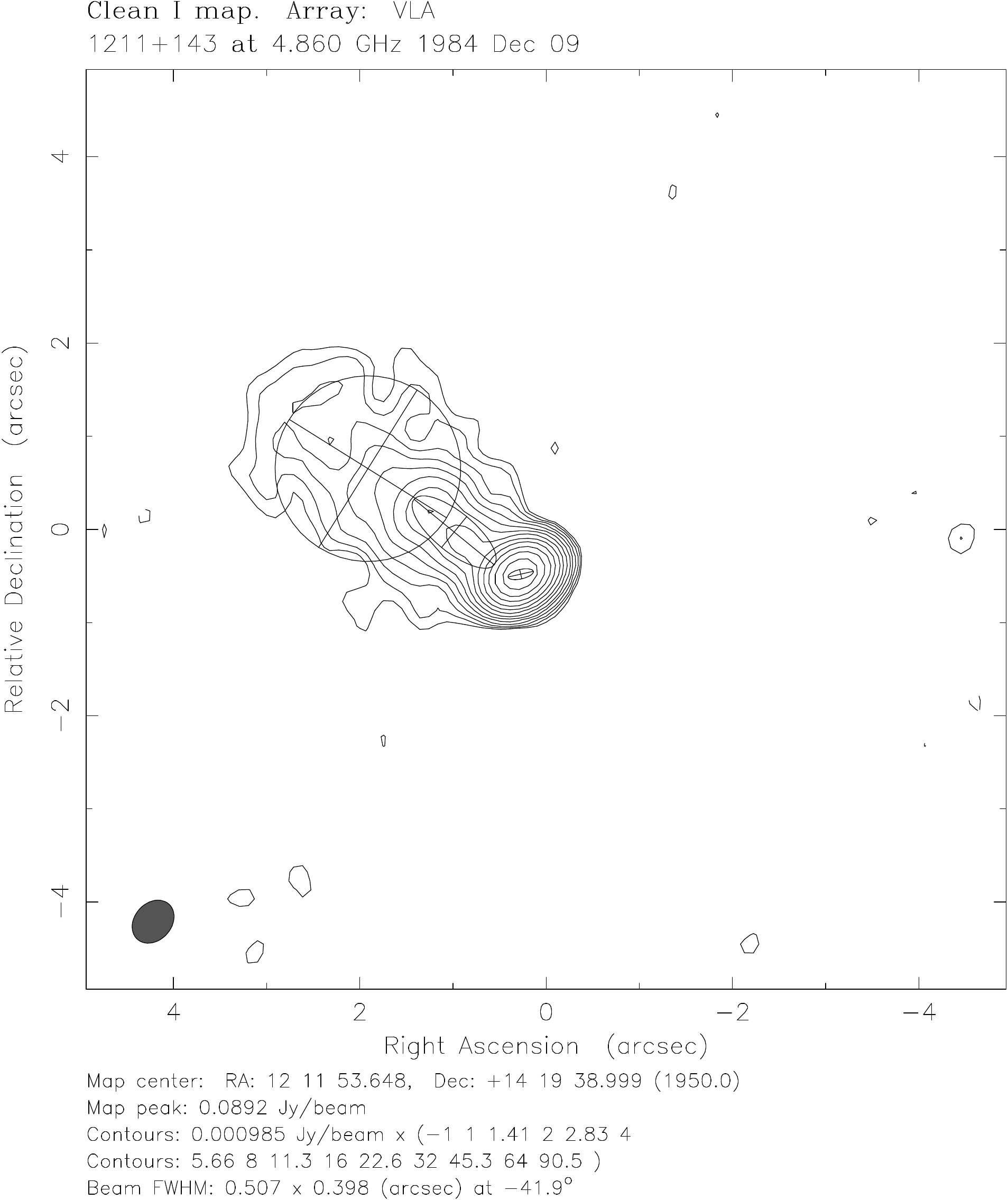}
	\includegraphics[width=0.95\columnwidth]{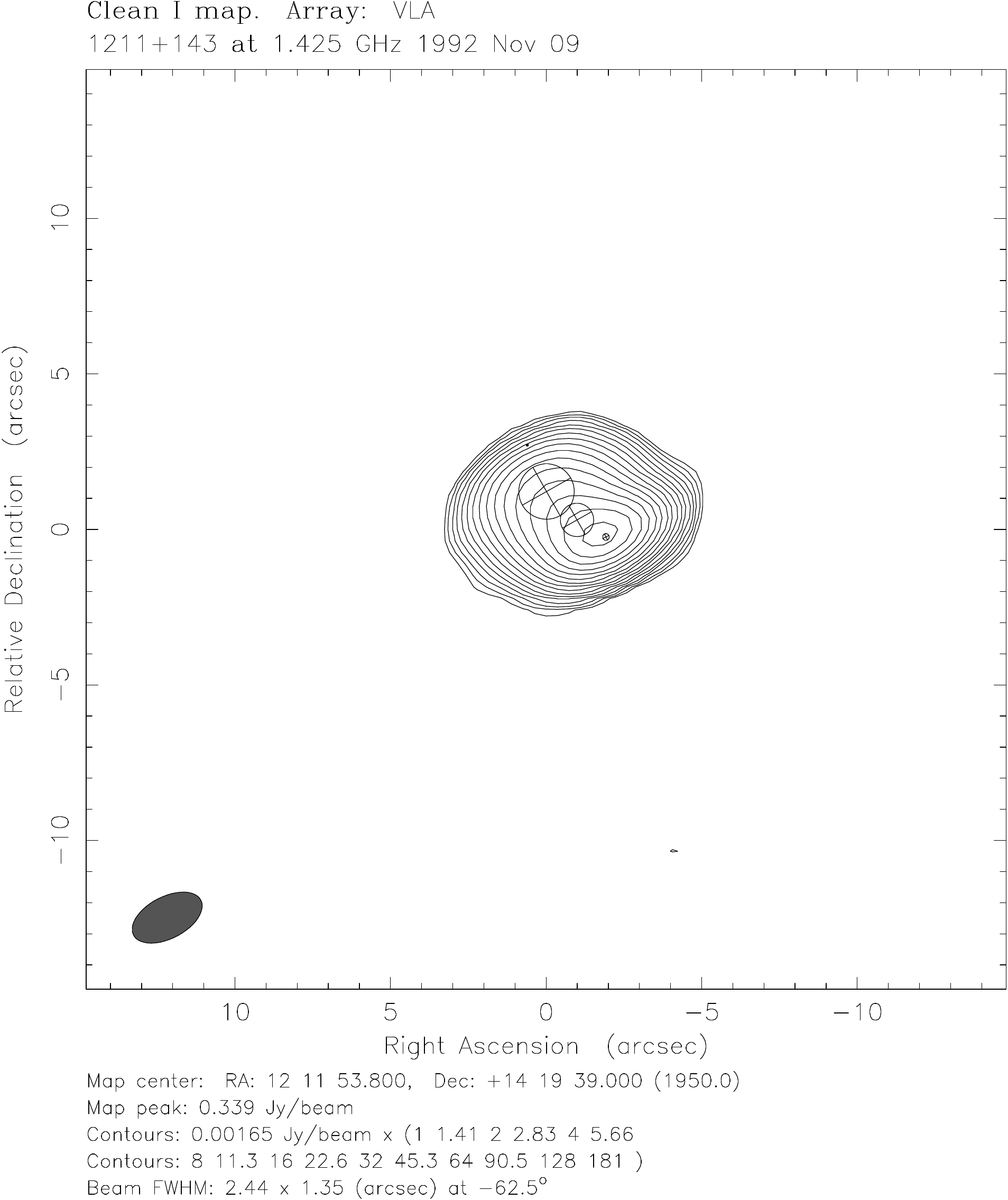}
	\includegraphics[width=0.95\columnwidth]{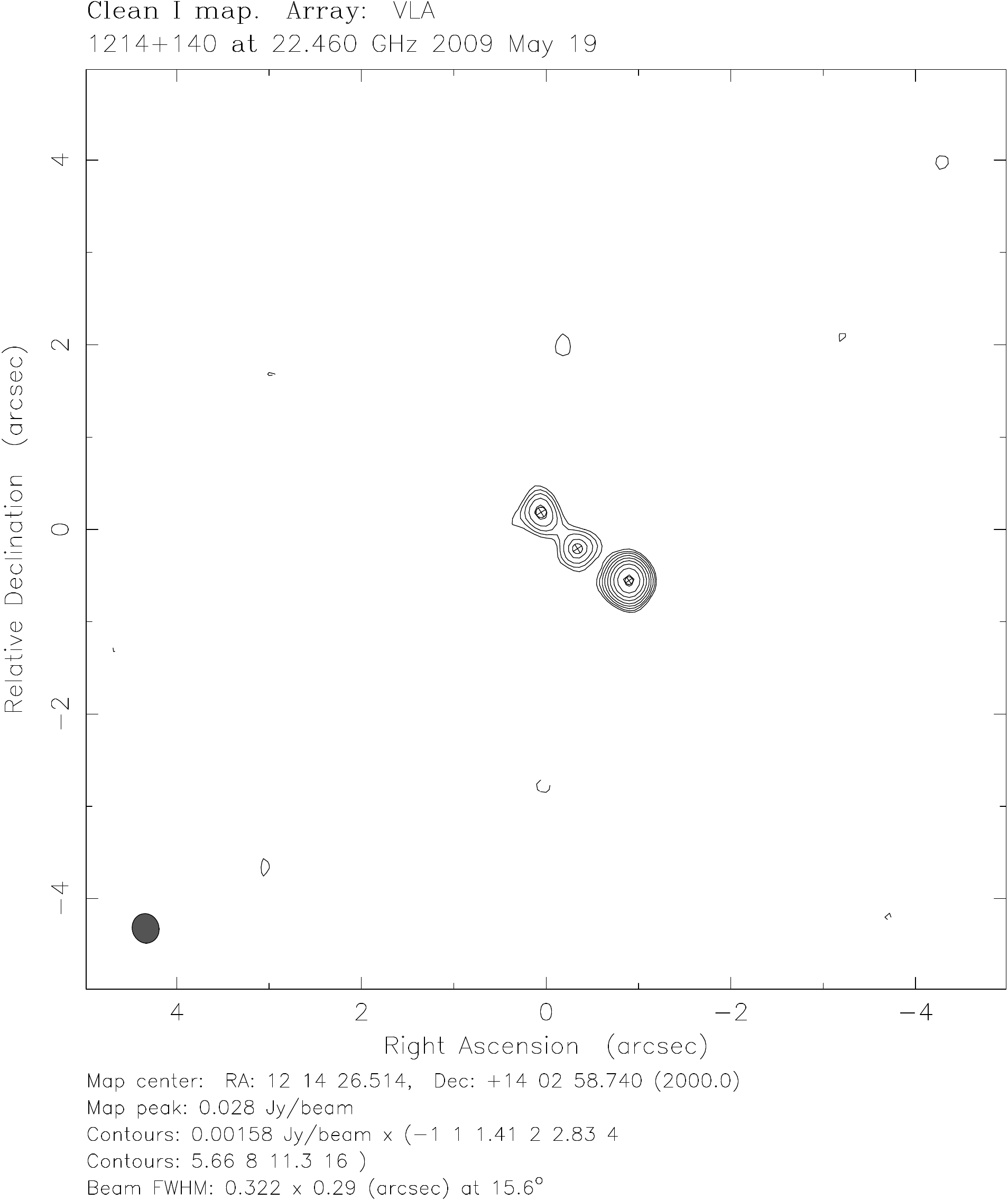}
	\includegraphics[width=0.95\columnwidth]{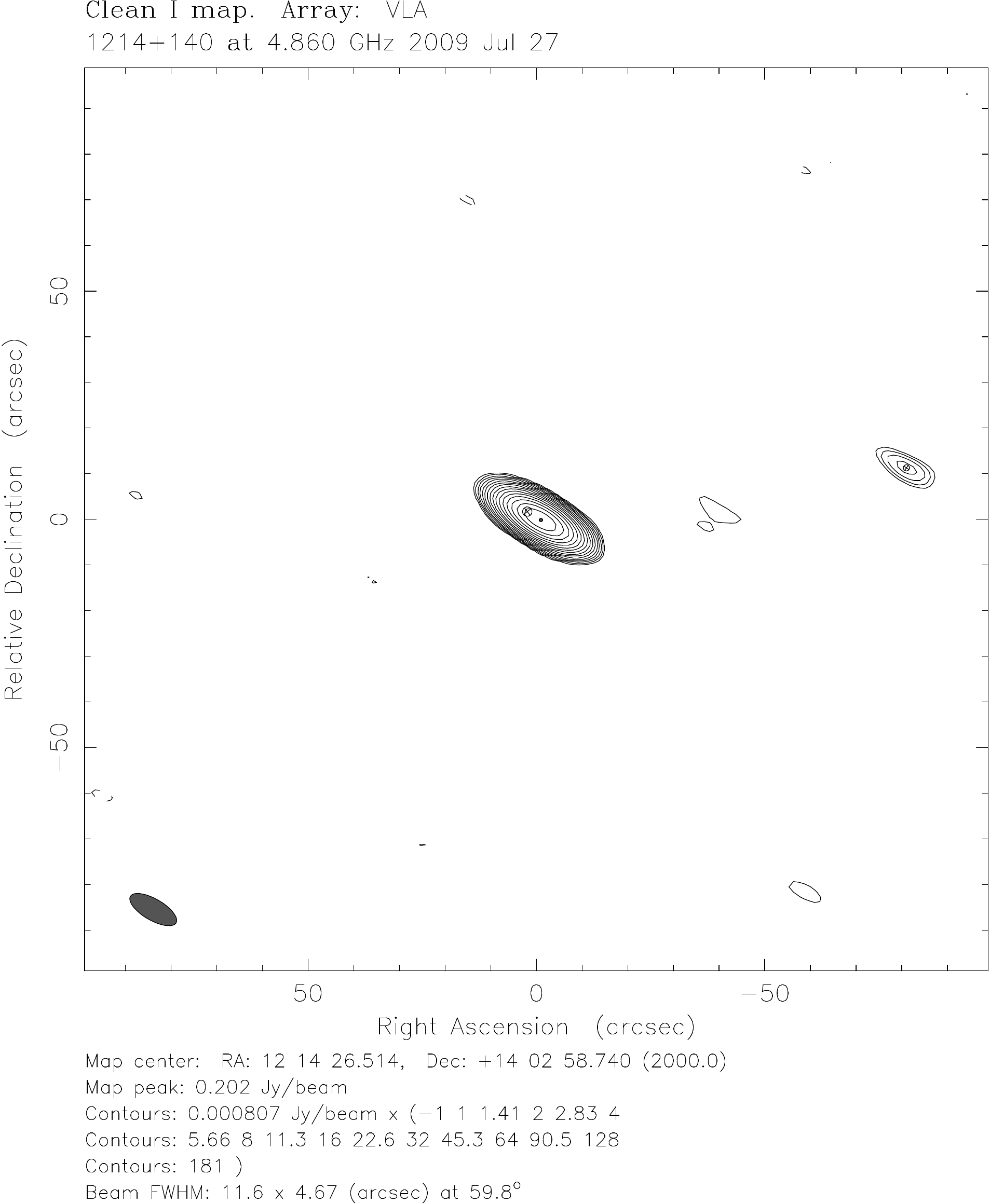}
	\caption{The archival VLA multi-band radio images of J1214+1402. The coordinate systems for upper and lower panels are B1950 and J2000, respectively.\label{fig:vla}}
\end{figure*}

\begin{figure}
	\centering
	\includegraphics[width=0.95\columnwidth]{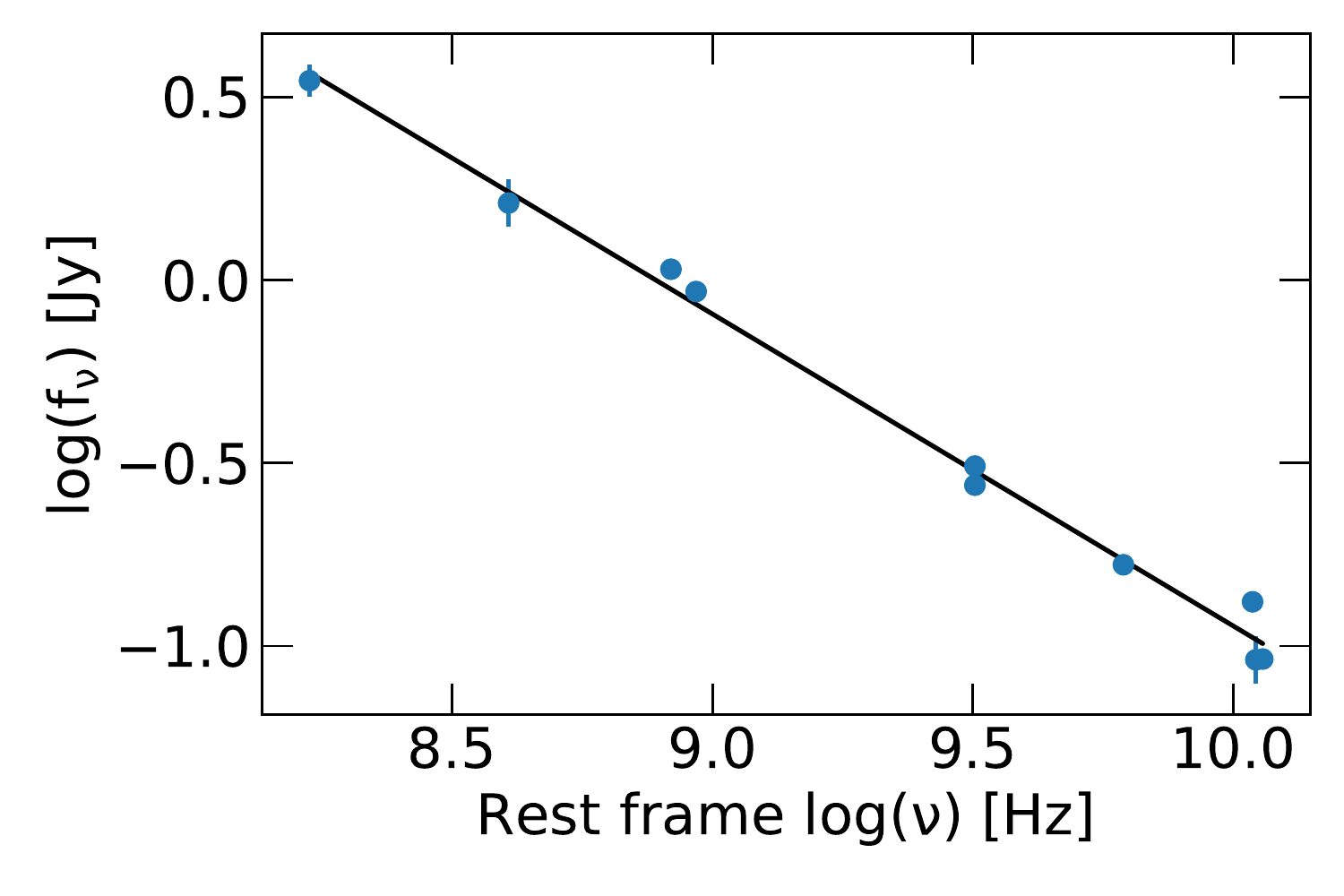}
	\caption{Radio continuum spectrum of J1214+1402.  \label{fig:sed_r}}
\end{figure}

The radio emission from the area around J1214+1402 was first discovered by the fourth Cambridge interferometer survey \cite[the source was identified as 4C +14.46, ][]{1967MmRAS..71...49G}, but the radio coordinates were often confused with SDSS J121428.25+140258.4 ($\alpha_{2000}=\rm 12^h14^m28.256^s$, $\delta_{2000}=\rm +14^d02^m58.48^s$, with separation to J1214+1402 about $25.2\arcsec$) for the low position accuracy and the low angular resolution in the following surveys \cite[e.g., ][]{1981MNRAS.194..693L, 1986ApJS...61....1B, 1991ApJS...75....1B, 1991ApJS...75.1011G, 1992ApJS...79..331W}. To obtain the accurate radio flux density of J1214+1402, we processed the historical Very Large Array (VLA) archived data with the Astronomical Image Processing System (AIPS) using standard procedure. There are four VLA continuum observations for J1214+1402, see details in Table \ref{tab:vla_info}. The standard flux density calibrator 3C 286 was used for every observation. Phase calibrators are listed in Table \ref{tab:vla_info}. After the calibration process, the imaging and model-fitting were performed in DIFMAP \cite[][]{1997ASPC..125...77S} and the final results and radio images are listed in Table \ref{tab:vla_info} and Figure \ref{fig:vla}. VLA images depict that the radio location is consistent with the optical coordinates of J1214+1402 and show no other radio object in the source-centered circle of 60 arc seconds radius. This indicates no radio emission from SDSS J121428.25+140258.4 and confirms that the radio emission should be associated to J1214+1402 \cite{1981MNRAS.194..693L, 1986ApJS...61....1B, 1991ApJS...75....1B, 1991ApJS...75.1011G, 1992ApJS...79..331W}. Based on all the published radio data, see Table \ref{tab:radio_info} and Figure \ref{fig:sed_r}, the slope of the radio continuum of J1214+1402 is about $\alpha = 0.85 $. 
Hence, J1214+1402 is a steep-spectrum radio quasar (SSRQ) rather than a flat-spectrum radio quasar (FSRQ), and therefore, its SED is less likely to be dominated by beamed jet.

\begin{table}
	\caption{VLA observations for J1214+1402}
	\label{tab:vla_info}
	\begin{threeparttable}
		\tiny
		\begin{tabular}{rccc|crrr}
			\hline
			\multicolumn{1}{c}{Project} & \multicolumn{1}{c}{Obs Date} & \multicolumn{1}{c}{Config.} & \multicolumn{1}{c}{Calibrator} & \multicolumn{1}{c}{Comp.} & \multicolumn{1}{c}{Flux} & \multicolumn{1}{c}{Distance} & \multicolumn{1}{c}{Theta} \\
			\multicolumn{4}{c}{ } & \multicolumn{1}{c}{ } & \multicolumn{1}{c}{$\rm Jy$} & \multicolumn{1}{c}{$\rm mas$} & \multicolumn{1}{c}{$\rm deg$} \\
			\multicolumn{1}{c}{(1)} & \multicolumn{1}{c}{(2)} & \multicolumn{1}{c}{(3)} & \multicolumn{1}{c}{(4)} & \multicolumn{1}{c}{(5)} & \multicolumn{1}{c}{(6)} & \multicolumn{1}{c}{(7)} & \multicolumn{1}{c}{(8)}\\	
			\hline
			\multicolumn{1}{l}{AH0167} & 1984 Dec 09    & A:C & 1147+245 & 0             & 0.103      & 0             & 0 \\
			&               &               &               & 1             & 0.049      & 845.316    & 57.643 \\
			&               &               &               & 2             & 0.044      & 1993.042     & 55.351 \\
			\multicolumn{1}{l}{ADHOC} & 1992 Nov 09    & A:L & 1219+285 & 0             & 0.283      & 0             & 0 \\
			&               &               &               & 1             & 0.186      & 1078.429    & 59.508 \\
			&               &               &               & 2             & 0.187      & 2403.922     & 52.565 \\
			\multicolumn{1}{l}{CALSUR} & 2009 May 19    & B:K & J1230+1223 & 0             & 0.030     & 0             & 0 \\
			&               &               &               & 1             & 0.006    & 656.465    & 58.202 \\
			&               &               &               & 2             & 0.008     & 921.694    & 50.276 \\
			\multicolumn{1}{l}{CALSUR} & 2009 Jul 27    & C:C & J1230+1223 & 0             & 0.197      & 0             & 0 \\
			&               &               &               & 1             & 0.011     & 3518.330    & 58.245 \\
			\hline
		\end{tabular}
		\begin{tablenotes}
			\footnotesize
			\item In this Table, Column (1): The NRAO observing project ID; Column (2): VLA observational date; Column (3): VLA Configuration: observing bands; Column (4): Phase calibrators; Column (5): Radio components, peak flux component marked as 0 which assumed to be a radio core, in this work; Column (6): model components flux density; Column (7-8): Components position relative to the peak component, and its position angle.
		\end{tablenotes}
	\end{threeparttable}
\end{table}%

\begin{table}
	\caption{Radio information of J1214+1402}
	\label{tab:radio_info}
	\begin{threeparttable}
		\begin{tabular}{rccr}
			\hline
			\multicolumn{1}{c}{Passband} & \multicolumn{1}{c}{Frequency} & \multicolumn{1}{c}{Flux} & \multicolumn{1}{c}{References} \\
			\multicolumn{1}{c}{$\rm MHz$} & \multicolumn{1}{c}{$\rm Hz$} & \multicolumn{1}{c}{$\rm Jy$} & \multicolumn{1}{c}{} \\
			\multicolumn{1}{c}{(1)} & \multicolumn{1}{c}{(2)} & \multicolumn{1}{c}{(3)} & \multicolumn{1}{c}{(4)} \\	
			\hline
			5000 & 5.00E+09 & 0.21 & \cite{1990PKS...C......0W} \\
			4850 & 4.85E+09 & $0.21\pm 0.03$ & \cite{1991ApJS...75.1011G} \\
			4775 & 4.75E+09 & $0.20$ & \cite{1991ApJS...75.1011G} \\
			2700 & 2.70E+09 & 0.38 & \cite{1990PKS...C......0W} \\
			1400 & 1.40E+09 & $0.71\pm0.02$ & \cite{1998AJ....115.1693C} \\
			1400 & 1.40E+09 & 0.63 & \cite{1992ApJS...79..331W} \\
			408 & 4.08E+08 & $2.12\pm0.11$ & \cite{1981MNRAS.194..693L} \\
			365 & 3.65E+08 & $2.44\pm0.06$  & \cite{1996AJ....111.1945D} \\
			178 & 1.78E+08 & $3.70\pm0.56$ & \cite{1967MmRAS..71...49G} \\
			73.8 & 7.38E+07 & $7.99\pm0.81$ & \cite{2007AJ....134.1245C} \\
			\hline
		\end{tabular}
		\begin{tablenotes}
			\footnotesize
			\item In this Table, Column (1): Observed passband; Column (2): Frequency; Column (3): Flux density; Column (4): References for radio flux in Column (3).
		\end{tablenotes}
	\end{threeparttable}
\end{table}%

\begin{figure}
	\centering
	\includegraphics[width=0.95\columnwidth]{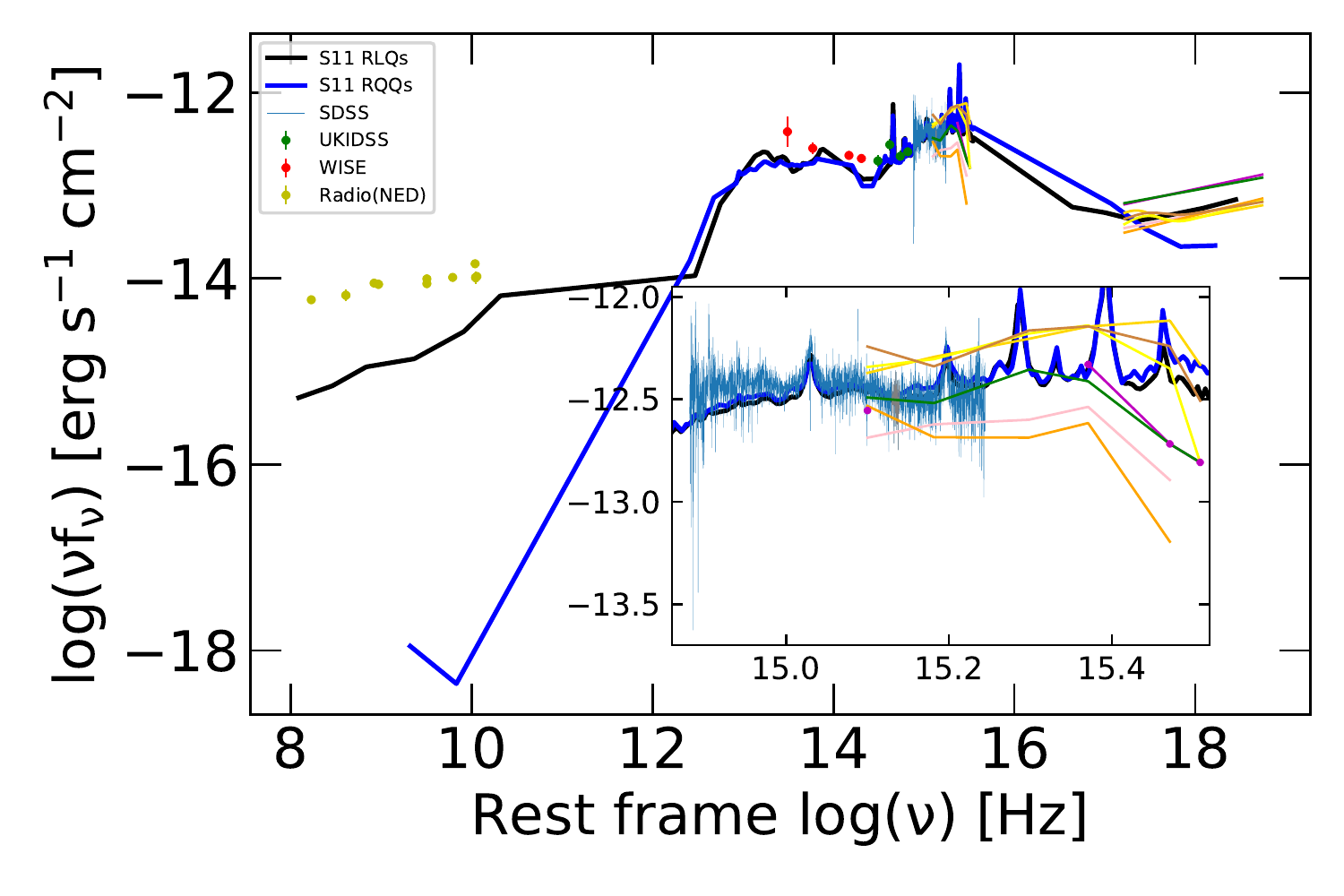}
	\caption{Broad band SED of J1214+1402. The thick black and blue solid lines are composite SEDs for RLQs and RQQs in S11, respectively. Colored dots show the different bands data in the upper-left corner, and the colored solid lines are the multi-epoch XMM-Newton optical/UV to X-ray spectra. The inset features the details of optical spectra. \label{fig:sed}}
\end{figure}

Blazars are one of the most enigmatic and rare classes of AGNs \cite[][]{2010ApJ...716...30A}. This population of AGNs is characterized by flat radio spectra, super-luminal motion, variable emission, and high polarization from radio to optical bands. Their emission is dominated by non-thermal radiation from radio to $\gamma$-ray bands \cite[][]{1995PASP..107..803U, 1998MNRAS.299..433F}. The J1214+1402 was identified to be a FSRQ candidate in \cite{2014ApJS..215...14D} using WISE data by $locus$ method \cite[see details in][]{2013ApJS..206...12D}. To verify, we plotted the broad band SED of J1214+1402 in Figure \ref{fig:sed} from radio to X-ray band using the SDSS optical data, multi-band radio data in Table \ref{tab:radio_info}, and NASA/IPAC Extragalactic Database (NED\footnote{\url{http://ned.ipac.caltech.edu/}}) archival Infrared data \cite[UKIDSS and WISE data were published by][respectively]{2010MNRAS.406.1583W, 2013wise.rept....1C}. 
Firstly, the SDSS optical spectrum was corrected for galactic extinction with the reddening map of \cite{1998ApJ...500..525S}. Then, the multi-band fluxes were also K corrected and redshift to the rest frame. After the data process of Section \ref{subsec:newton}, the XMM-Newton galactic extinction corrected optical/UV and the best-fitted X-ray spectra were added to Figure \ref{fig:sed} for demonstration. 
For comparison, the composite SEDs of RLQs and RQQs \cite[][hereafter S11]{2011ApJS..196....2S} were also plotted with their 2200 \AA~ flux fixed to the mean flux density within $50$ \AA ~ around the rest-frame $2200$ \AA ~ in the SDSS spectrum of J1214+1402. 
Similar to the composite SEDs of RLQs and RQQs, the broad band SED of J1214+1402 manifests a prominent big blue bump (BBB) and the infrared bump. This implies that the infrared-to-optical flux of J1214+1402 may not be dominated by the non-thermal processes of the beamed jet, but the thermal emission of dust torus and accretion disk \citep[e.g., ][]{1993ARA&A..31..473A, 2015ARA&A..53..365N}.

In $\gamma$-ray band, we cross match J1214+1402 coordinates with Fermi LAT 8-year Source Catalog \cite[4FGL\footnote{\url{https://fermi.gsfc.nasa.gov/ssc/data/access/lat/8yr_catalog/}},][]{2020ApJS..247...33A} within $5\deg$ \cite[based on][the Large Area Telescope on the Fermi Gamma-Ray Space Telescope has angular resolution from a few arc minutes for the highest-energy photons to $ 3.5 $ degrees at $ 100~\rm MeV $]{2009ApJ...697.1071A}. 
It gives out 14 corresponding sources within $5$ degree separation, but there are no $ \gamma $-ray counterparts within positional uncertainty.
This suggests that J1214+1402 has no $ \gamma $-ray emission detected by Fermi-LAT. 

Overall, the broadband emission of source J1214+1402 indicates that it is a misaligned radio source, and emission may not be significantly affected by the jet beaming effect.

\subsection{Swift}
\label{subsec:swift}

\begin{figure}
	\centering
	\includegraphics[width=1\columnwidth]{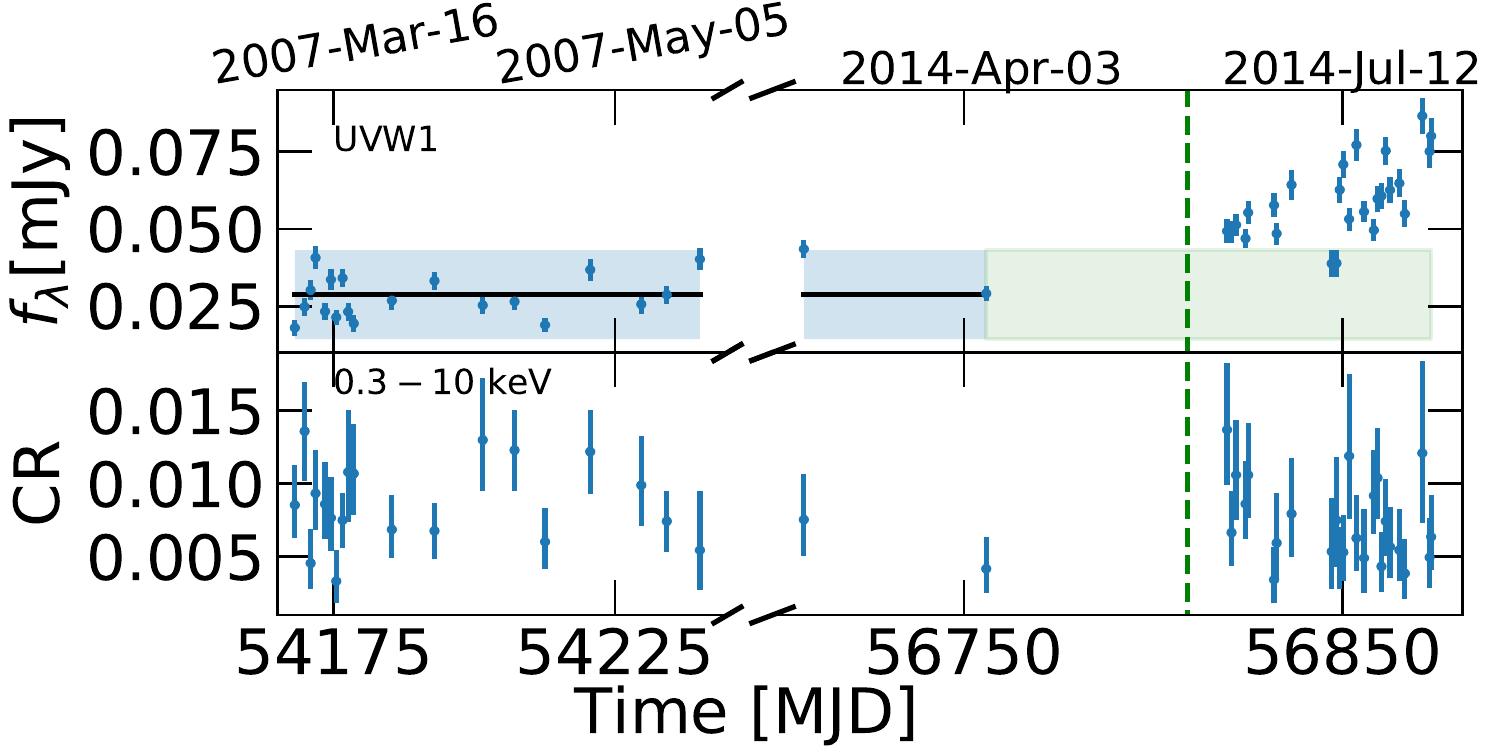}
	\caption{The Swift light curves of J1214+1402 for UVW1 instrument (upper panel) and $0.3-10$ keV X-ray band (lower panel). The black solid line and blue-shaded region correspond to the mean and $2\sigma$ uncertainty of the photometric flux observed by the Swift/UVW1 instrument before 2014-Apr-08. For a better understanding of the variation in the full-time range, the blue-shaded region is extended as the green-shaded region. The green dashed line represents the time position of 2014 June 01. }
   \label{fig:swift}
\end{figure}

J1214+1402 has been observed by Swift observatory 72 times from 2007 to 2018. In order to inspect the Swift UVOT and XRT data, we generate the X-ray light curve in the energy range of $0.3-10.0~\rm keV$ using the XRT on-line data analysis tools\footnote{\url{https://www.swift.ac.uk/user_objects/}} \citep{2009MNRAS.397.1177E}, and extract UVOT flux with the on-line UVOT interactive analysis tool\footnote{\url{https://swift.ssdc.asi.it/}}. In total, there are 46 observations that have well-detected data of J1214+1402 for both XRT and UVOT ($ UVW1 $) instruments. The details of the X-ray count rate and $ UVW1 $ fluxes are displayed in Table \ref{tab:swift} and the light curves are shown in Figure \ref{fig:swift}.

\begin{table}
	\centering
	\caption{The Swift observations of J1214+1402}
	\label{tab:swift}
	\begin{threeparttable}
	\begin{tabular}{l|ccc}
	\hline
	\multicolumn{1}{c}{Obs ID} & \multicolumn{1}{c}{Obs date} & \multicolumn{1}{c}{Rate} & \multicolumn{1}{c}{$f_{\rm \lambda, uvw1}$} \\
	\multicolumn{1}{c}{ } & \multicolumn{1}{c}{ } & \multicolumn{1}{c}{$0.01~\rm cts/s$} & \multicolumn{1}{c}{$0.01~\rm mJy$} \\
	\multicolumn{1}{c}{(1)} & \multicolumn{1}{c}{(2)} & \multicolumn{1}{c}{(3)} & \multicolumn{1}{c}{(4)} \\
	\hline
	00030904002   & 2007 Mar 09    & $ 0.85^{+0.27}_{-0.23} $ & $ 1.80\pm{0.25} $ \\
	00030904003   & 2007 Mar 10    & $ 1.36^{+0.34}_{-0.34} $ & $ 2.48\pm{0.29} $ \\
	00030904004   & 2007 Mar 11    & $ 0.45^{+0.23}_{-0.18} $ & $ 3.02\pm{0.32} $ \\
	00030904005   & 2007 Mar 12    & $ 0.93^{+0.30}_{-0.25} $ & $ 4.07\pm{0.38} $ \\
	00030904007   & 2007 Mar 14    & $ 0.86^{+0.29}_{-0.24} $ & $ 2.33\pm{0.27} $ \\
	00030904008   & 2007 Mar 15    & $ 0.77^{+0.28}_{-0.23} $ & $ 3.36\pm{0.35} $ \\
	00030904009   & 2007 Mar 15    & $ 0.33^{+0.21}_{-0.15} $ & $ 2.14\pm{0.25} $ \\
	00030904010   & 2007 Mar 17    & $ 0.75^{+0.19}_{-0.19} $ & $ 3.41\pm{0.28} $ \\
	00030904011   & 2007 Mar 18    & $ 1.08^{+0.43}_{-0.34} $ & $ 2.32\pm{0.28} $ \\
	00030904012   & 2007 Mar 19    & $ 1.07^{+0.34}_{-0.28} $ & $ 1.94\pm{0.26} $ \\
	00030904013   & 2007 Mar 26    & $ 0.69^{+0.24}_{-0.20} $ & $ 2.68\pm{0.29} $ \\
	00030904014   & 2007 Apr 02    & $ 0.68^{+0.19}_{-0.19} $ & $ 3.32\pm{0.30} $ \\
	00030904016   & 2007 Apr 11    & $ 1.30^{+0.43}_{-0.35} $ & $ 2.53\pm{0.29} $ \\
	00030904017   & 2007 Apr 17    & $ 1.23^{+0.28}_{-0.28} $ & $ 2.65\pm{0.29} $ \\
	00030904018   & 2007 Apr 22    & $ 0.60^{+0.23}_{-0.19} $ & $ 1.89\pm{0.24} $ \\
	00030904019   & 2007 Apr 30    & $ 1.22^{+0.29}_{-0.29} $ & $ 3.68\pm{0.35} $ \\
	00030904021   & 2007 May 09    & $ 0.99^{+0.34}_{-0.28} $ & $ 2.56\pm{0.31} $ \\
	00030904022   & 2007 May 14    & $ 0.74^{+0.21}_{-0.21} $ & $ 2.87\pm{0.29} $ \\
	00030904023   & 2007 May 20    & $ 0.54^{+0.41}_{-0.27} $ & $ 4.02\pm{0.35} $ \\
	00080664001   & 2014 Feb 19    & $ 0.75^{+0.32}_{-0.25} $ & $ 4.35\pm{0.28} $ \\
	00080664002   & 2014 Apr 08    & $ 0.42^{+0.22}_{-0.17} $ & $ 2.91\pm{0.24} $ \\
	00030904030   & 2014 Jun 11    & $ 1.37^{+0.46}_{-0.38} $ & $ 4.93\pm{0.38} $ \\
	00030904031   & 2014 Jun 12    & $ 0.67^{+0.28}_{-0.23} $ & $ 4.91\pm{0.35} $ \\
	00030904032   & 2014 Jun 13    & $ 1.06^{+0.38}_{-0.31} $ & $ 5.13\pm{0.36} $ \\
	00030904035   & 2014 Jun 16    & $ 0.86^{+0.30}_{-0.24} $ & $ 4.69\pm{0.32} $ \\
	00030904036   & 2014 Jun 17    & $ 1.06^{+0.36}_{-0.29} $ & $ 5.53\pm{0.37} $ \\
	00030904041   & 2014 Jun 23    & $ 0.34^{+0.23}_{-0.16} $ & $ 5.77\pm{0.38} $ \\
	00030904042   & 2014 Jun 24    & $ 0.59^{+0.34}_{-0.25} $ & $ 4.85\pm{0.35} $ \\
	00030904044   & 2014 Jun 28    & $ 0.79^{+0.39}_{-0.30} $ & $ 6.43\pm{0.48} $ \\
	00030904045   & 2014 Jul 09    & $ 0.54^{+0.36}_{-0.26} $ & $ 3.88\pm{0.44} $ \\
	00030904046   & 2014 Jul 10    & $ 0.75^{+0.43}_{-0.32} $ & $ 3.89\pm{0.43} $ \\
	00030904047   & 2014 Jul 11    & $ 0.46^{+0.24}_{-0.18} $ & $ 6.27\pm{0.42} $ \\
	00030904048   & 2014 Jul 12    & $ 0.53^{+0.25}_{-0.20} $ & $ 7.09\pm{0.43} $ \\
	00030904049   & 2014 Jul 13    & $ 1.19^{+0.56}_{-0.43} $ & $ 5.32\pm{0.38} $ \\
	00030904051   & 2014 Jul 15    & $ 0.63^{+0.29}_{-0.23} $ & $ 7.72\pm{0.51} $ \\
	00030904053   & 2014 Jul 17    & $ 0.49^{+0.34}_{-0.24} $ & $ 5.56\pm{0.34} $ \\
	00030904056   & 2014 Jul 20    & $ 0.92^{+0.32}_{-0.26} $ & $ 4.96\pm{0.36} $ \\
	00030904057   & 2014 Jul 21    & $ 1.04^{+0.34}_{-0.29} $ & $ 5.98\pm{0.42} $ \\
	00030904058   & 2014 Jul 22    & $ 0.43^{+0.23}_{-0.18} $ & $ 6.07\pm{0.41} $ \\
	00030904059   & 2014 Jul 23    & $ 0.74^{+0.29}_{-0.24} $ & $ 7.53\pm{0.46} $ \\
	00030904060   & 2014 Jul 24    & $ 0.57^{+0.27}_{-0.22} $ & $ 6.26\pm{0.43} $ \\
	00030904062   & 2014 Jul 26    & $ 0.54^{+0.28}_{-0.21} $ & $ 6.48\pm{0.46} $ \\
	00030904064   & 2014 Jul 28    & $ 0.39^{+0.23}_{-0.17} $ & $ 5.49\pm{0.44} $ \\
	00030904066   & 2014 Aug 02    & $ 1.21^{+0.63}_{-0.48} $ & $ 8.66\pm{0.58} $ \\
	00030904067   & 2014 Aug 03    & $ 0.50^{+0.27}_{-0.21} $ & $ 7.51\pm{0.54} $ \\
	00030904068   & 2014 Aug 04    & $ 0.64^{+0.29}_{-0.23} $ & $ 8.01\pm{0.58} $ \\	
	\hline
	\end{tabular}
	\begin{tablenotes}
		\footnotesize
		\item In this Table, Column (1): Swift observational ID; Column (2): Swift observational date; Column (3): X-ray count rate in $0.3-10.0~\rm keV$; Column (4): UV flux in UVW1 band.
	\end{tablenotes}
	\end{threeparttable}
\end{table}%

As displayed in Figure \ref{fig:swift}, the light curve of J1214+1402 shows two prominent optical/UV flux states, a low state before 2014 April 8 and a high state after 2014 June 11. The mean value of the X-ray count rate for these two time ranges are $0.008\pm 0.003~ \rm cts~s^{-1}$ and $0.007\pm 0.003~ \rm cts~s^{-1}$ for the former and latter, respectively. The intrinsic variation amplitude (the excess variance $\sigma_{\rm rms}$) for these two time ranges is less than 0, where $\sigma_{\rm rms}$ is defined as $\sigma_{\rm rms}^2=\frac{1}{N-1} \sum(X_i-\Bar{X})^2-\frac{1}{N}\sum \sigma_i^2 $ \cite[see, e.g.,][]{2018ApJ...868...58K}, with $N$ being the number of X-ray measurements, $X_i$ the X-ray count rates, $\Bar{X}$ the average count rates, and $\sigma_i$ the uncertainty of each observation. 
Hence, the Swift X-ray light curve shows no significant count rate variation between these two optical/UV states.  

To study the Swift X-ray spectra, with the XSELECT v2.4k package in HEASOFT software, we combine all Swift X-ray Level 2 event files into two parts (before and after 2014 June) based on $UVW1$ flux. 
The source and background spectra are extracted from a source-centered circle of a 15-pixel radius and a nearby source-free region of a 60-pixel radius circle around the object, respectively. 
In the end, two combined X-ray spectra are rebinned with 15 counts per bin, and fitted by X{\footnotesize SPEC} v12.9 with $\chi^2$ statistical methods. 
We find that both X-ray spectra can be fitted with an absorbed power-law model ($phabs*zphabs*powerlaw$) with galactic {H\,{\footnotesize I}} column density fixed to $\rm2.60\times10^{20}~ cm^{-2}$ \citep{2005A&A...440..775K}. 
The best-fitting power-law parameters are $\Gamma=1.63^{+0.20}_{-0.19}$, $Norm=4.57^{+0.59}_{-0.29}\times10^{-5}~\rm photons/keV/cm^2/s$ and $\Gamma=1.77^{+0.35}_{-0.30}$, $Norm=4.49^{+1.24}_{-0.98}\times10^{-5}~\rm photons/keV/cm^2/s$ for low and high optical/UV state, respectively. The corresponding fluxes at rest frame $2~\rm keV$ are $\log \nu f_{\nu,2\rm keV}=-13.16^{+0.07}_{-0.04} ~\rm erg\ cm^{-2}\ s^{-1}$ and $\log \nu f_{\nu,2\rm keV}=-13.16^{+0.14}_{-0.11} ~\rm erg\ cm^{-2}\ s^{-1}$.  

\subsection{XMM-Newton}
\label{subsec:newton}

\begin{figure*}
	\includegraphics[width=1.4\columnwidth]{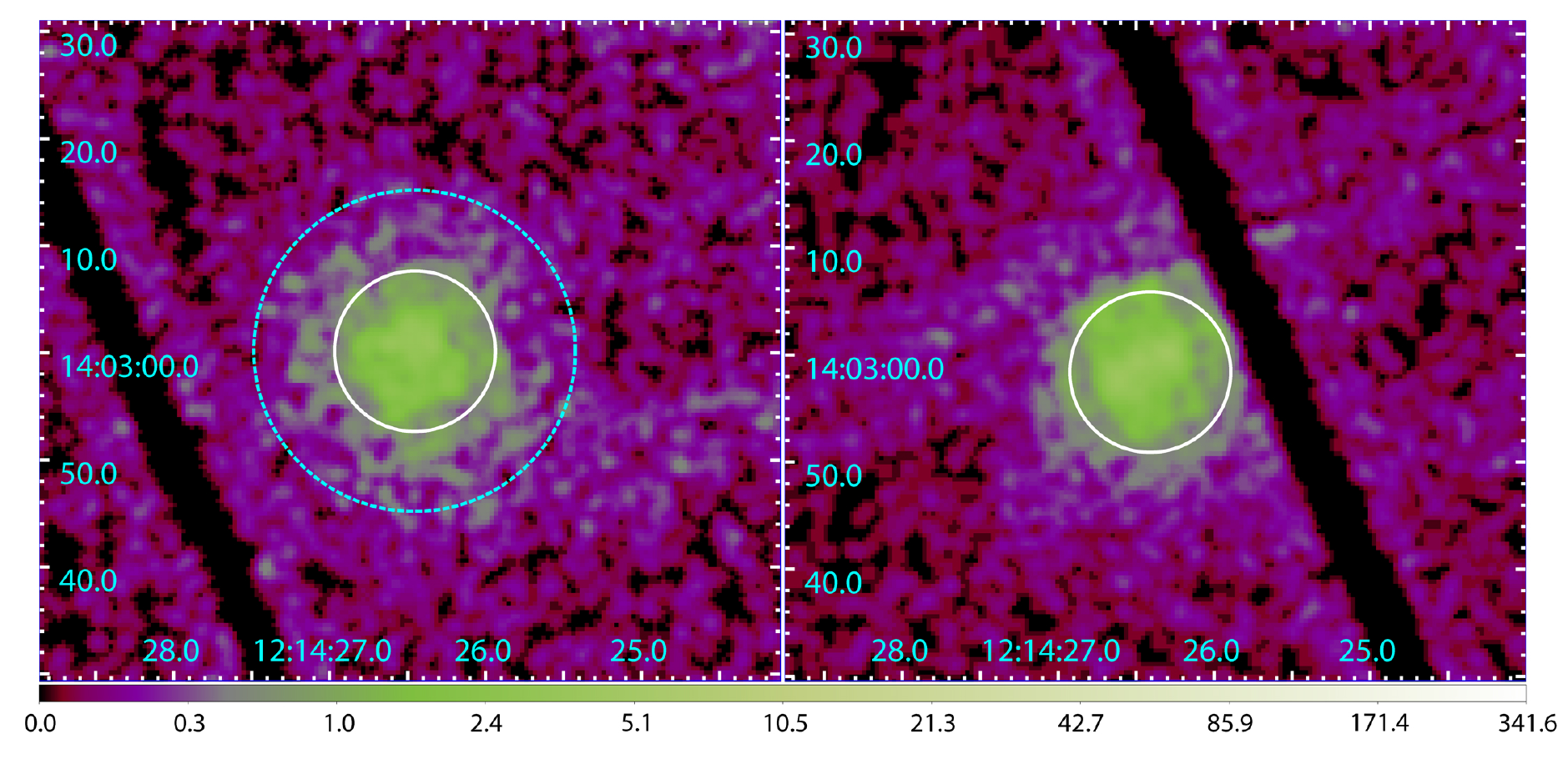}
	\caption{Examples of XMM-Newton images of J1214+1402 smoothed with $3\sigma$ kernel, left for 0502050101, right for 0745110401. The applied core extraction regions with the 15\arcsec and 7.5\arcsec radius are represented with cyan dashed circles and white solid circles, respectively. }
	\label{fig:xmm_image}
\end{figure*}

The J1214+1402 was observed 12 times by XMM-Newton, as shown in Table \ref{tab:xmm_info}. Only three observations (ID: 0208020101, 0502050101, and 0502050201) have MOS detection.
We only used pn data for X-ray spectra analysis, except for 0208020101, and processed all data with XMM-Newton Scientific Analysis Software (SAS) using SAS cookbook\footnote{\url{https://heasarc.gsfc.nasa.gov/docs/xmm/abc/}} step by step. The pn (or MOS) data-sets were reprocessed with $epproc$ (or $emproc$) scripts in SAS-15.0.0, and were filtered with standard filters: PATTERN in the range of 0 to 4 (or 0 to 12), energy in the range of $0.2-15.0~\rm keV$ (or $0.2-12.0~\rm keV$), and \#XMMEA\_EP (or \#XMMEA\_EM).
The large flare time interval was filtered out with light curve checking. The observation 0112610201 has a large background flare during all exposure time, and so has no available X-ray photon data. 
We extracted J1214+1402 source spectra with $evselect$ from a source-centered circle of radius $15\arcsec$ and/or $7.5\arcsec$, and the background spectra from a nearby source-free circle region with radius $40\arcsec$. 
If the source is near the edge of a pn CCD chip, we only used the source-centered circle of radius $7.5\arcsec$. 
Figure \ref{fig:xmm_image} displays the images of J1214+1402 for two different XMM-Newton observations. As shown in the left panel (0502050101), we extracted the source X-ray spectrum from the circle of $15\arcsec$ and $7.5\arcsec$ radius. However, for the right panel (0745110401), the spectrum can only be extracted from $7.5\arcsec$ radius circle as the source is close to the CCD edge. 
In the cases that J1214+1402 is near the edge of pn CCD chip with a distance less than $7.5\arcsec$, we extracted the X-ray spectrum only if MOS data are available (e.g., 0208020101), or do not analyse the X-ray data for others which have only pn data (as for 0745110101, 0745110201, 0745110301, and 0745110501). 
After careful inspections, we were left with only seven available X-ray observational data, on which the detailed spectral analysis can be performed (see Table \ref{tab:xmm_xfit}), and their spectra are not affected by pile-up as checked by $epatplot$. 
The redistribution matrix files (RMF) and ancillary response files (ARF) were created with $rmfgen$ and $arfgen$ scripts, respectively. 
Finally, all spectra were rebinned with a minimum of 15 counts for background-subtracted spectral channel and oversampling the intrinsic energy resolution by a factor of no larger than three.

\begin{table}
	\caption{XMM-Newton observations for J1214+1402}
	\label{tab:xmm_info}
	\begin{threeparttable}
	\begin{tabular}{ccrc}
	\hline
	\multicolumn{1}{c}{XMM ID} & \multicolumn{1}{c}{Obs Date} & \multicolumn{1}{c}{Duration [s]} & \multicolumn{1}{c}{OM bands} \\
	\multicolumn{1}{c}{(1)} & \multicolumn{1}{c}{(2)} & \multicolumn{1}{c}{(3)} & \multicolumn{1}{c}{(4)} \\	
	\hline
	0112610101    & 2001 Jun 15 & 55,738 & \multicolumn{1}{c}{UVW1/M2/W2} \\
	0112610201    & 2001 Jun 15 & 5,369  & \multicolumn{1}{c}{V} \\
	0208020101    & 2004 Jun 21 & 60,033 & \multicolumn{1}{c}{ALL} \\
	0502050101    & 2007 Dec 21 & 64,869 & \multicolumn{1}{c}{ALL} \\
	0502050201    & 2007 Dec 23 & 51,069 & \multicolumn{1}{c}{ALL} \\
	0745110101    & 2014 Jun 02 & 87,000 & \multicolumn{1}{c}{ALL} \\
	0745110201    & 2014 Jun 16 & 104,000 & \multicolumn{1}{c}{ALL} \\
	0745110301    & 2014 Jun 20 & 102,500 & \multicolumn{1}{c}{ALL} \\
	0745110401    & 2014 Jun 23 & 100,000 & \multicolumn{1}{c}{ALL} \\
	0745110501    & 2014 Jun 25 & 58,000 & \multicolumn{1}{c}{ALL} \\
	0745110601    & 2014 Jun 30 & 95,300 & \multicolumn{1}{c}{ALL} \\
	0745110701    & 2014 Jul 07 & 99,000 & \multicolumn{1}{c}{ALL} \\
	\hline
	\end{tabular}
	\begin{tablenotes}
	\footnotesize
	\item In this Table, Column (1): XMM-Newton observational ID; Column (2): XMM-Newton observational date; Column (3): Time duration in $\rm s$; Column (4): OM optical/UV bands, `ALL' for all six OM bands, V/B/U/UVW1/UVM2/UVW2.
	\end{tablenotes}
	\end{threeparttable}
\end{table}%

\begin{figure*}
	\includegraphics[width=0.65\columnwidth,angle=-90]{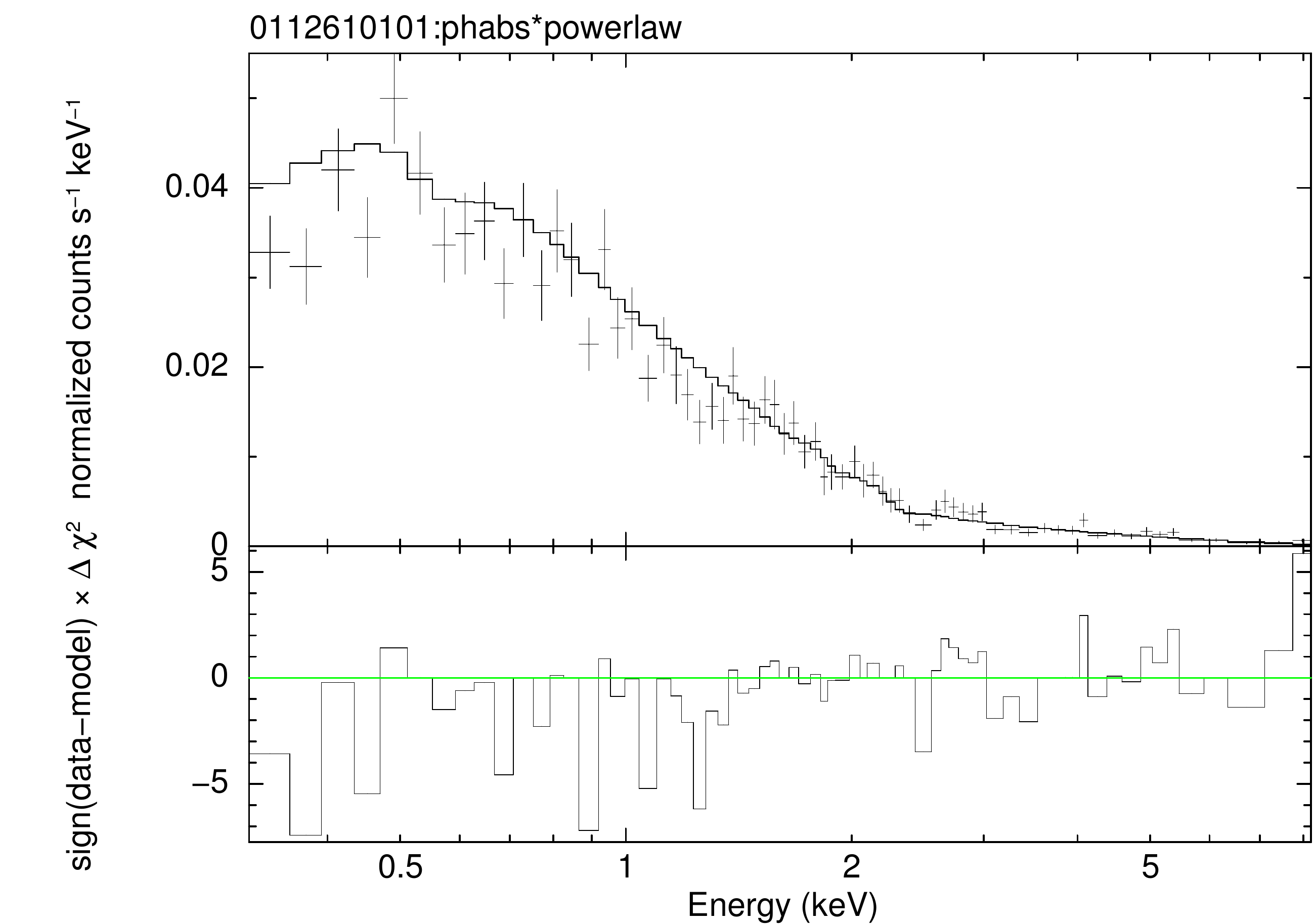}
	\includegraphics[width=0.65\columnwidth,angle=-90]{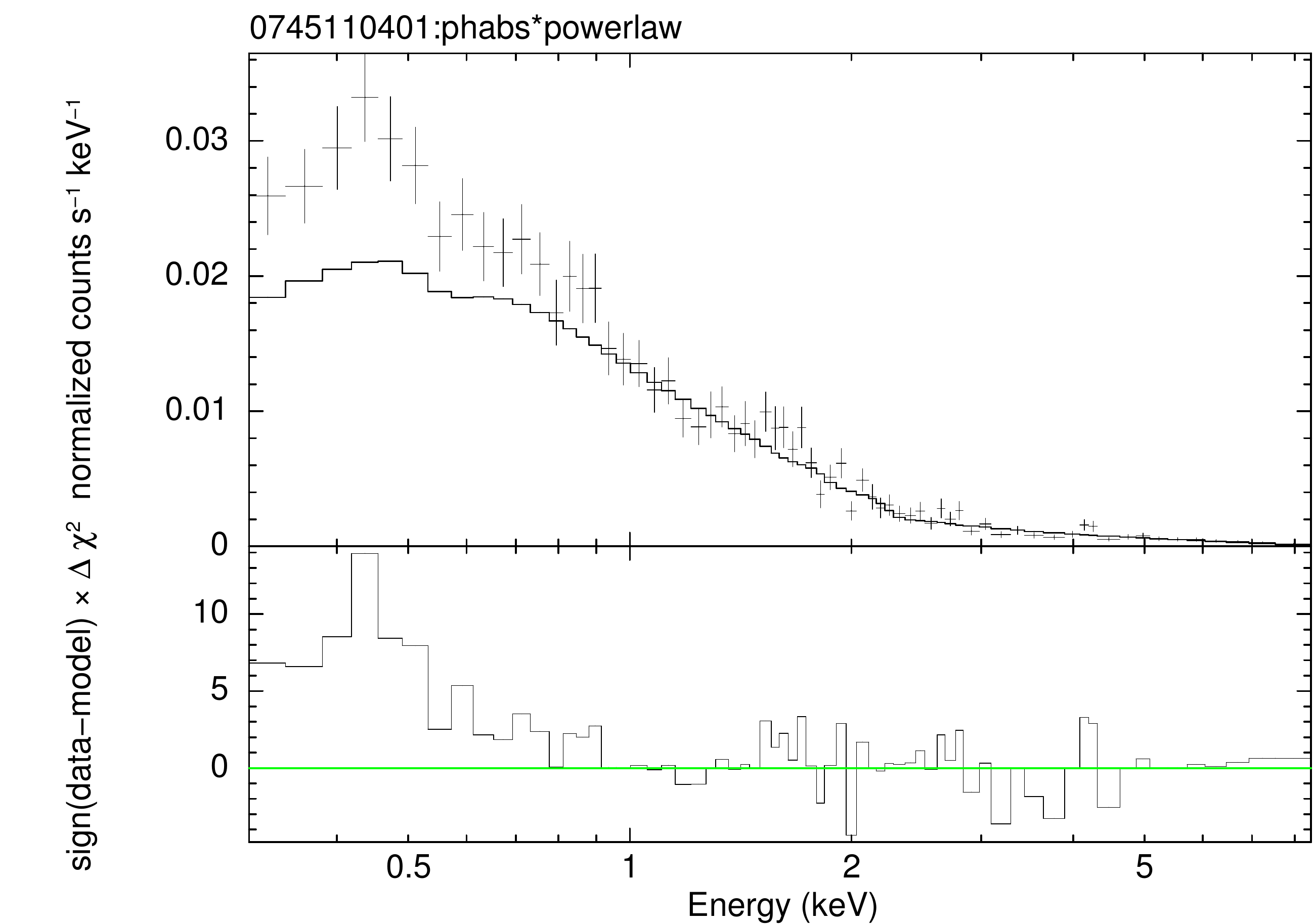}
	\includegraphics[width=0.65\columnwidth,angle=-90]{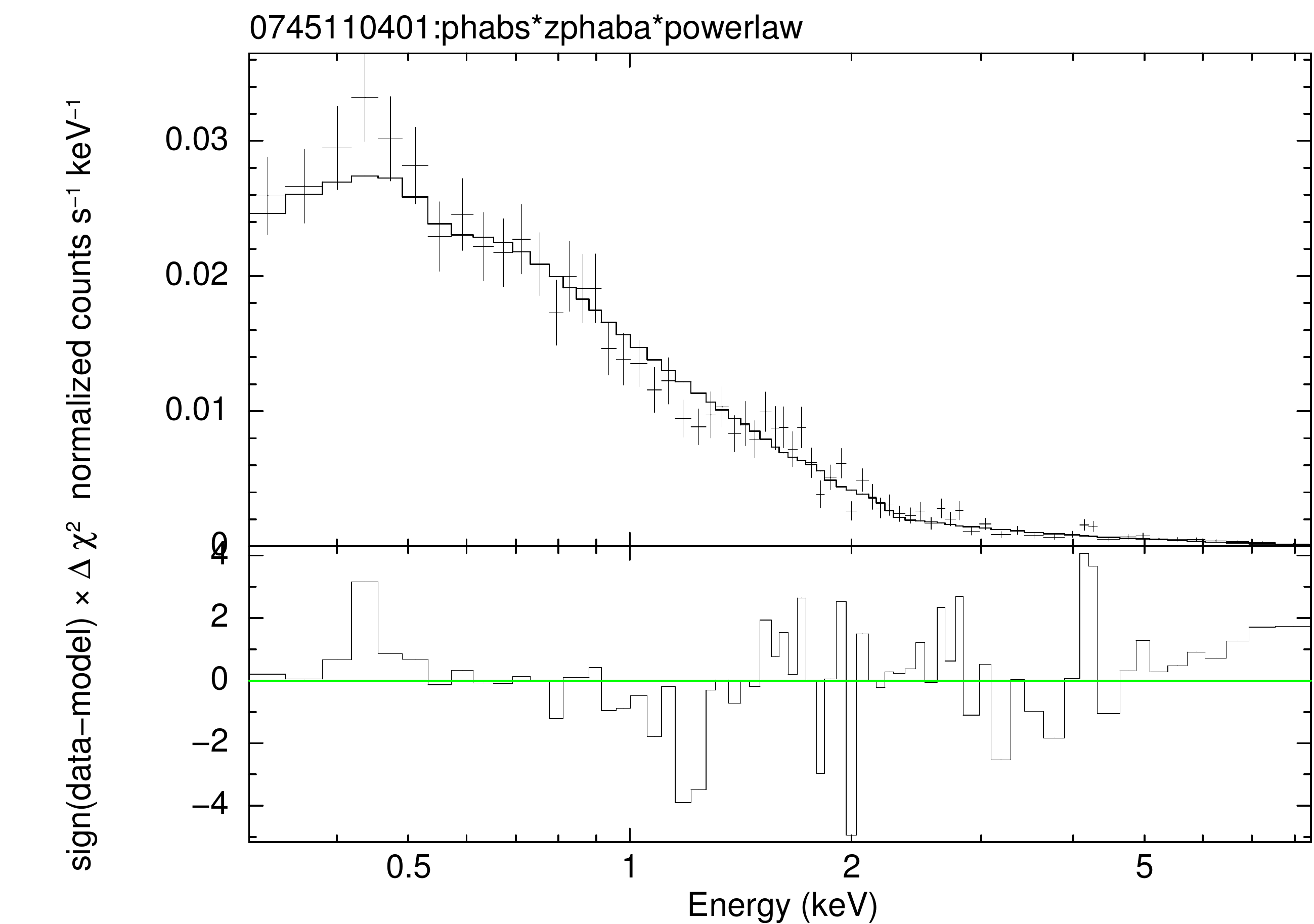}
	\includegraphics[width=0.65\columnwidth,angle=-90]{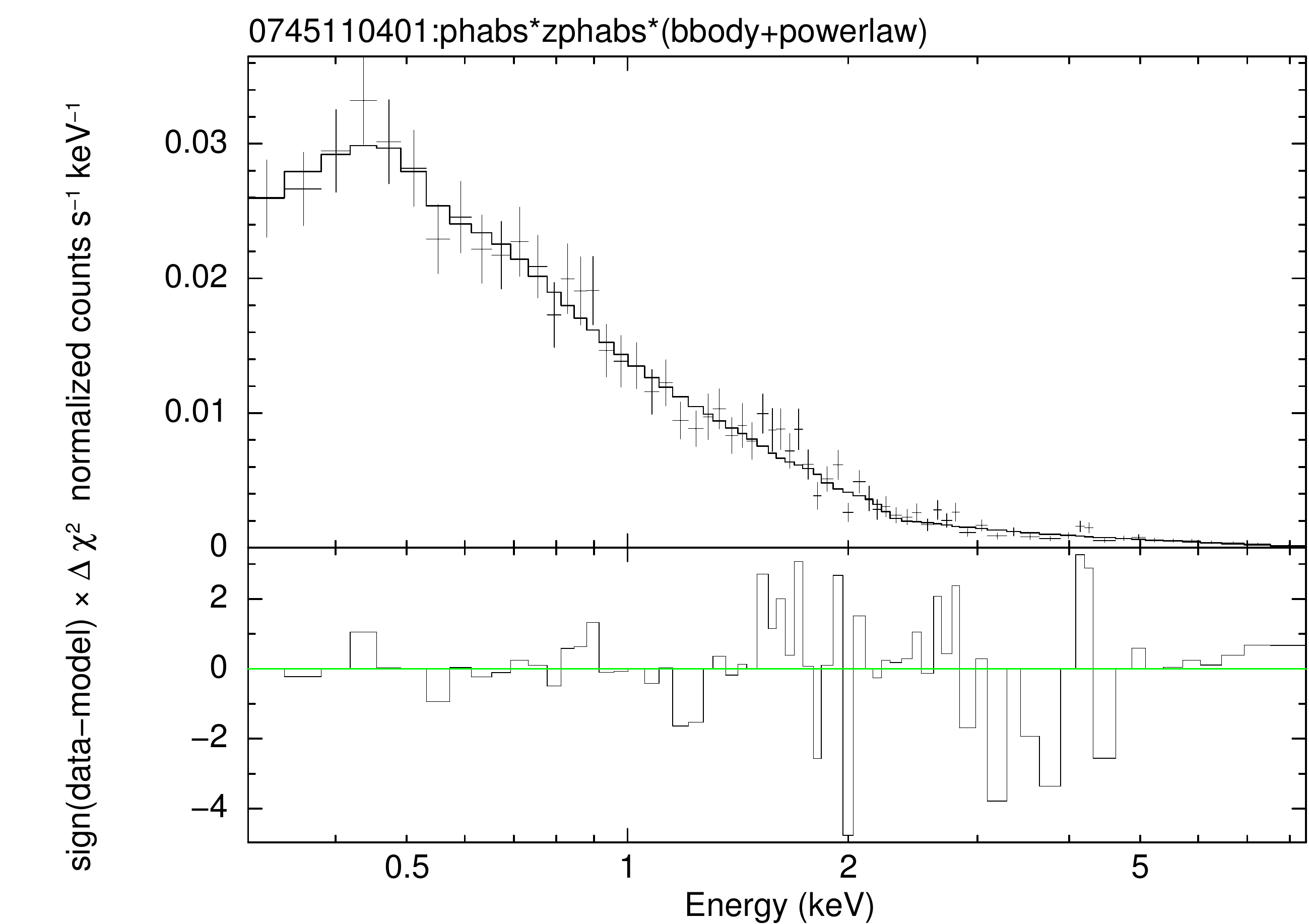}
	\caption{Examples of X-ray spectral fitting. \textit{Upper panels}: spectra fitted with the galactic absorbed power-law in $2.0-10.0~\rm keV$ and the resulting residuals after extending to the whole spectrum (the left for 0112610101, and the right for 0745110401). \textit{Bottom panels}: spectra fitted with different models in $0.3-10.0~\rm keV$ and the resulting residuals for 0745110401 (the left for absorbed power-law model, and the right for absorption corrected black-body plus power-law model).}
	\label{fig:xmm_spec}
\end{figure*}

All XMM-Newton X-ray spectra were fitted by X{\footnotesize SPEC} v12.9 with $\chi^2$ statistical methods. Following the method to search for soft X-ray excess in \cite{2021MNRAS.507.3572X} and \cite{2021RAA....21....4Z}, we firstly fitted X-ray spectra in $2.0-10.0~\rm keV$ with a galactic absorbed \cite[$\rm nH=2.60\times10^{20}~cm^{-2}$,][]{2005A&A...440..775K} single power-law model ($phabs*powerlaw$).
The fitted power-law was then extended to $0.3-2.0~\rm keV$, from which the residual between the data and the power-law model was checked. We found that three spectra have prominent soft X-ray excess components below $1~\rm keV$. As shown in the upper panels of Figure \ref{fig:xmm_spec}, compared to 0112610101, the spectrum of 0745110401 has significant soft X-ray excess. We fitted the whole spectrum of these three data with galactic absorption corrected black-body plus power-law model ($phabs*zphabs*(bbody+powerlaw)$) and also single power-law model ($phabs*zphabs*powerlaw$). Compared to the single power-law model, as shown in the bottom panels of Figure \ref{fig:xmm_spec}, the black-body plus power-law model gives better fitting, especially at the soft and hard energy tail. 
The other four data were fitted by a single absorbed power-law model ($phabs*zphabs*powerlaw$) on the whole spectrum. The best fitting results are listed in Table \ref{tab:xmm_xfit}. To calculate the error for each parameter in spectral fitting with 90\% confidence level, we use the `error' command in X{\footnotesize SPEC}. 

\begin{table*}
	\caption{XMM-Newton X-ray spectral fit results}
	\label{tab:xmm_xfit}
	\begin{threeparttable}
		\begin{tabular}{llrrllllccc}
			\hline
			\multicolumn{1}{c}{XMM ID} & \multicolumn{1}{c}{Obs Date} & \multicolumn{1}{c}{Radius} & \multicolumn{1}{c}{nH} & \multicolumn{1}{c}{kT} & \multicolumn{1}{c}{Bbo.norm} & \multicolumn{1}{c}{$\Gamma$} & \multicolumn{1}{c}{Pow.norm} & \multicolumn{1}{c}{$\chi^2_{Red}$} & \multicolumn{1}{c}{D.O.F.} & \multicolumn{1}{c}{$\log \nu f_{\nu,2\rm keV}$} \\
			\multicolumn{1}{c}{ } & \multicolumn{1}{c}{ } & \multicolumn{1}{c}{$\arcsec$} & \multicolumn{1}{c}{$ 10^{22} \rm cm^{-2}$} & \multicolumn{1}{c}{$\rm keV$} & \multicolumn{1}{c}{$\rm E-07$} & \multicolumn{1}{c}{ } & \multicolumn{1}{c}{$\rm E-05$} & \multicolumn{1}{c}{ } & \multicolumn{1}{c}{ } & \multicolumn{1}{c}{$\rm erg~cm^{-2}~s^{-1}$} \\
			\multicolumn{1}{c}{(1)} & \multicolumn{1}{c}{(2)} & \multicolumn{1}{c}{(3)} & \multicolumn{1}{c}{(4)} & \multicolumn{1}{c}{(5)} & \multicolumn{1}{c}{(6)} & \multicolumn{1}{c}{(7)} & \multicolumn{1}{c}{(8)} & \multicolumn{1}{c}{(9)} & \multicolumn{1}{c}{(10)} & \multicolumn{1}{c}{(11)} \\	
			\hline
			0112610101    & 2001 Jun 15  & 7.5  & $ \leq 0.07 $      &   &   & $ 1.80_{-0.05}^{+0.08}$  & $ 3.06_{-0.11}^{+0.20} $ & 1.02 & 69 & $ -13.32^{+0.03}_{-0.02} $ \\
			&    & 15.0  & $ \leq 0.03 $      &   &   & $ 1.79_{-0.04}^{+0.05}$  & $ 5.03_{-0.15}^{+0.19} $ & 1.05 & 88 & $ -13.11^{+0.02}_{-0.02} $ \\
			0208020101    & 2004 Jun 21  & 15.0   & $ \leq 0.01 $      &   &   & $ 1.82_{-0.09}^{+0.10}$  & $ 5.00_{-0.31}^{+0.38} $ & 1.04 & 51 & $ -13.11^{+0.04}_{-0.03} $ \\
			0502050101    & 2007 Dec 21  & 7.5   & $ \leq 0.14 $      &   &   & $ 1.80_{-0.10}^{+0.12}$  & $ 1.67_{-0.12}^{+0.15} $ & 0.93 & 52 & $ -13.58^{+0.05}_{-0.04} $ \\
			&    & 15.0  & $ \leq 0.10 $      &   &   & $ 1.81_{-0.07}^{+0.09}$  & $ 2.73_{-0.15}^{+0.18} $ & 0.93 & 70 & $ -13.37^{+0.03}_{-0.03} $ \\
			0502050201    & 2007 Dec 23  & 7.5   & $ \leq 0.10 $      &   &   & $ 1.73_{-0.08}^{+0.15}$  & $ 1.68_{-0.10}^{+0.21} $ & 0.85  & 42 & $ -13.59^{+0.06}_{-0.03} $ \\
			&    & 15.0  & $ \leq 0.07 $      &   &   & $ 1.76_{-0.07}^{+0.09}$  & $ 2.59_{-0.13}^{+0.18} $ & 1.11 & 53 & $ -13.40^{+0.04}_{-0.03} $ \\
			0745110401    & 2014 Jun 23  & 7.5   & $ \leq 0.35 $      & $ 0.10_{-0.03}^{+0.03}$ & $2.05_{-1.23}^{+11.85}$ & $ 1.82_{-0.11}^{+0.13}$  & $ 1.58_{-0.17}^{+0.21} $ & 0.90  & 68 & $ -13.58^{+0.29}_{-0.12} $ \\
			0745110601    & 2014 Jun 30 & 7.5    & $ \leq 0.26 $     & $ 0.12_{-0.04}^{+0.02}$ & $1.57_{-0.67}^{+5.46}$ & $ 1.74_{-0.11}^{+0.13}$  & $ 1.48_{-0.16}^{+0.21} $ & 0.82  & 70 & $ -13.57^{+0.33}_{-0.14} $ \\
			0745110701    & 2014 Jul 07 & 7.5    & $ \leq 0.01 $      & $ 0.13_{-0.13}^{+0.05}$ & $0.99_{-0.85}^{+0.80}$ & $ 1.83_{-0.12}^{+0.15}$  & $ 1.79_{-0.21}^{+0.27} $ & 1.02  & 73 & $ -13.51^{+0.14}_{-0.21} $ \\
			\hline
		\end{tabular}
		\begin{tablenotes}
			\footnotesize
			\item In this Table, Column (1): XMM-Newton observational ID; Column (2): XMM-Newton observational date; Column (3): The radius of source extraction region; Column (4): Intrinsic hydrogen column density; Column (5$ - $6): The temperature and normalization (in $ 10^{-7}~\rm photons~ keV^{-1}~ cm^{-2}~ s^{-1} $ at $ 1~\rm keV $) parameters with $1~\sigma$ errors of black body component; Column (7): The power-law photon index and $1~\sigma$ errors; Column (8): The normalization and $1~\sigma$ errors of power-law component in $ 10^{-5}~\rm photons~ keV^{-1}~ cm^{-2}~ s^{-1} $ at $ 1~\rm keV $ ; Column (9$ - $10): Reduced $ \chi^2 $ and degree of freedom; Column (11): Rest frame $2~\rm keV$ flux calculated from the correlate spectrum fitting and $1~\sigma$ errors.
		\end{tablenotes}
	\end{threeparttable}
\end{table*}%

From the XMM-Newton images of J1214+1402 in Figure \ref{fig:xmm_image}, we know that the source-centered circle with $15\arcsec$ radius covers almost all source X-ray photons (about $95\%$), while the circle with $7.5\arcsec$ radius covers only about $60\%$ photons. 
Comparing the X-ray spectra extracted from different regions in the same data, we found these two spectra have almost the same photon index but have different power-law normalization, as shown in Table \ref{tab:xmm_xfit}. This indicates that the hard and soft X-rays have a similar point-spread function (PSF) in the XMM-Newton pn detector. The X-ray flux at rest frame $2~\rm keV$ extracted from the region of $15\arcsec$ radius circles are higher than that of extracted from $7.5\arcsec$ radius, with an average value of $0.20~\rm dex$.  
Therefore, the X-ray emission from the circle of $7.5\arcsec$ radius can be corrected by $0.20~\rm dex$. 

Optical/UV OM image data has been fully processed by the XMM-SAS Pipeline. We directly use the pipeline results to calculate the optical flux in each band. To gain photon count rate or Magnitude, the interactive tool $omsource$ has been used. All corrected OM count rates are shown in Table \ref{tab:xmm_omcts}. 

With the optical/UV observations, we also calculated the Eddington ratio for each XMM-Newton OM data. First, XMM-Newton OM results were converted to flux with standard conversion factors\footnote{\url{https://xmm-tools.cosmos.esa.int/external/xmm_user_support/documentation/sas_usg/USG/}}. 
Then, optical/UV flux was corrected for galactic extinction using the reddening map of \cite{1998ApJ...500..525S}. And the luminosity at $3000$ \AA ~($L_{3000}$) was estimated by OM $V$ band flux, assuming SDSS optical continuum slope $\alpha = -0.94\pm 0.02$ \cite[$f_{\rm \lambda} \varpropto \lambda^{\alpha}$, as in][]{2020ApJS..249...17R}. In the end, the bolometric luminosity is calculated from monochromatic luminosity using the same correction factor as in \cite{2011ApJS..194...45S}, $L_{\rm bol}=5.15\times L_{3000}$. The Eddington ratio $L_{\rm bol}/L_{\rm Edd}$ is listed in Table \ref{tab:xmm_omcts}, where Eddington luminosity $L_{\rm Edd}$ was calculated with virial black hole masses as $L_{\rm Edd}=1.25\times 10^{38}(M_{\rm BH}/{\rm M_\odot})\, \rm erg\,s^{-1}$.

\begin{table*}
	\caption{XMM-Newton Optical/UV count rate }
	\label{tab:xmm_omcts}
	\begin{threeparttable}
	\begin{tabular}{llllllllcc}
		\hline
		\multicolumn{1}{c}{XMM ID} & \multicolumn{1}{c}{Obs Date} & \multicolumn{1}{c}{V} & \multicolumn{1}{c}{B} & \multicolumn{1}{c}{U} & \multicolumn{1}{c}{UVW1} & \multicolumn{1}{c}{UVM2} & \multicolumn{1}{c}{UVW2} & \multicolumn{1}{c}{$\log \nu f_{\nu,2500}$} & \multicolumn{1}{c}{Edd. ratio} \\ 
		\multicolumn{1}{c}{(1)} & \multicolumn{1}{c}{(2)} & \multicolumn{1}{c}{(3)} & \multicolumn{1}{c}{(4)} & \multicolumn{1}{c}{(5)} & \multicolumn{1}{c}{(6)} & \multicolumn{1}{c}{(7)} & \multicolumn{1}{c}{(8)} & \multicolumn{1}{c}{(9)} & \multicolumn{1}{c}{(10)} \\	
		\hline
		0112610101    & 2001 Jun 15    & $ 0.19\pm{0.08} $    &            &         & $ 0.29\pm{0.03} $    & $ 0.03\pm{0.02} $    & $ 0.01\pm{0.02} $ & $-12.59\pm 0.18$ & 0.045 \\
		0208020101    & 2004 Jun 21    & $ 0.22\pm{0.03} $    & $ 0.47\pm{0.04} $    & $ 0.58\pm{0.02} $    & $ 0.24\pm{0.01} $    & $ 0.03\pm{0.01} $    & $ 0.01\pm{0.01}$ & $-12.53\pm 0.06$ & 0.052 \\
		0502050101    & 2007 Dec 21    & $ 0.14\pm{0.07} $    & $ 0.37\pm{0.08} $    & $ 0.33\pm{0.05} $    & $ 0.18\pm{0.02} $    & $ 0.02\pm{0.01} $    & $ \le 0.01$ & $-12.72\pm 0.22$ & 0.033 \\
		0502050201    & 2007 Dec 23    & $ 0.20\pm{0.07} $    & $ 0.32\pm{0.09} $    & $ 0.27\pm{0.05} $    & $ 0.15\pm{0.02} $    & $ 0.01\pm{0.01} $    & $ \le 0.01$ & $-12.57\pm 0.15$ & 0.047 \\
		0745110101    & 2014 Jun 02    & $ 0.34\pm{0.05} $    & $ 0.83\pm{0.06} $    & $ 0.87\pm{0.04} $    & $ 0.41\pm{0.02} $    & $ 0.07\pm{0.01} $    & $ 0.01\pm{0.01}$ & $-12.34\pm 0.06$ & 0.081 \\
		0745110201    & 2014 Jun 16    & $ 0.37\pm{0.05} $    & $ 0.72\pm{0.06} $    & $ 0.75\pm{0.04} $    & $ 0.43\pm{0.02} $    & $ 0.09\pm{0.01} $    & $ \le 0.01 $ & $-12.30\pm 0.06$ & 0.088 \\
		0745110301    & 2014 Jun 20    & $ 0.36\pm{0.06} $    & $ 0.84\pm{0.07} $    & $ 0.88\pm{0.04} $    & $ 0.45\pm{0.02} $    & $ 0.08\pm{0.01} $    & $ 0.03\pm{0.01}$ & $-12.31\pm 0.07$ & 0.085 \\
		0745110401    & 2014 Jun 23    & $ 0.29\pm{0.06} $    & $ 0.79\pm{0.07} $    & $ 0.82\pm{0.04} $    & $ 0.44\pm{0.02} $    & $ 0.12\pm{0.02} $    & $ 0.03\pm{0.01}$ & $-12.41\pm 0.09$ & 0.069 \\
		0745110501    & 2014 Jun 25    & $ 0.48\pm{0.06} $    & $ 0.81\pm{0.07} $    & $ 0.80\pm{0.04} $    & $ 0.44\pm{0.02} $    & $ 0.07\pm{0.01} $    & $ 0.04\pm{0.01}$ & $-12.19\pm 0.05$ & 0.114 \\
		0745110601    & 2014 Jun 30    & $ 0.31\pm{0.06} $    & $ 0.77\pm{0.07} $    & $ 0.87\pm{0.04} $    & $ 0.45\pm{0.02} $    & $ 0.07\pm{0.01} $    & $ 0.01\pm{0.01}$ & $-12.38\pm 0.08$ & 0.073 \\
		0745110701    & 2014 Jul 07    & $ 0.39\pm{0.05} $    & $ 0.71\pm{0.06} $    & $ 0.90\pm{0.04} $    & $ 0.45\pm{0.02} $    & $ 0.09\pm{0.01} $    & $ 0.02\pm{0.01}$ & $-12.28\pm 0.06$ & 0.093 \\
		\hline
	\end{tabular}
	\begin{tablenotes}
	\footnotesize
	\item In this Table, Column (1): XMM-Newton observational ID; Column (2): XMM-Newton observational date; Column (3$-$8): V/B/U/UVW1/UVM2/UVW2 bands count rate [$\rm cts/s$] and $1~\sigma$ errors, with a note 0112610101 has no V band data and the correlate value is obtained from 0112610201 which observed in the same day of 0112610101; Column (9): Rest frame $2500$ \AA ~flux [$\rm erg~cm^{-2}~s^{-1}$] and $1~\sigma$ errors; Column (10): Eddington ratio.
	\end{tablenotes}
	\end{threeparttable}
\end{table*}%

\begin{figure}
	\includegraphics[width=1.0\columnwidth]{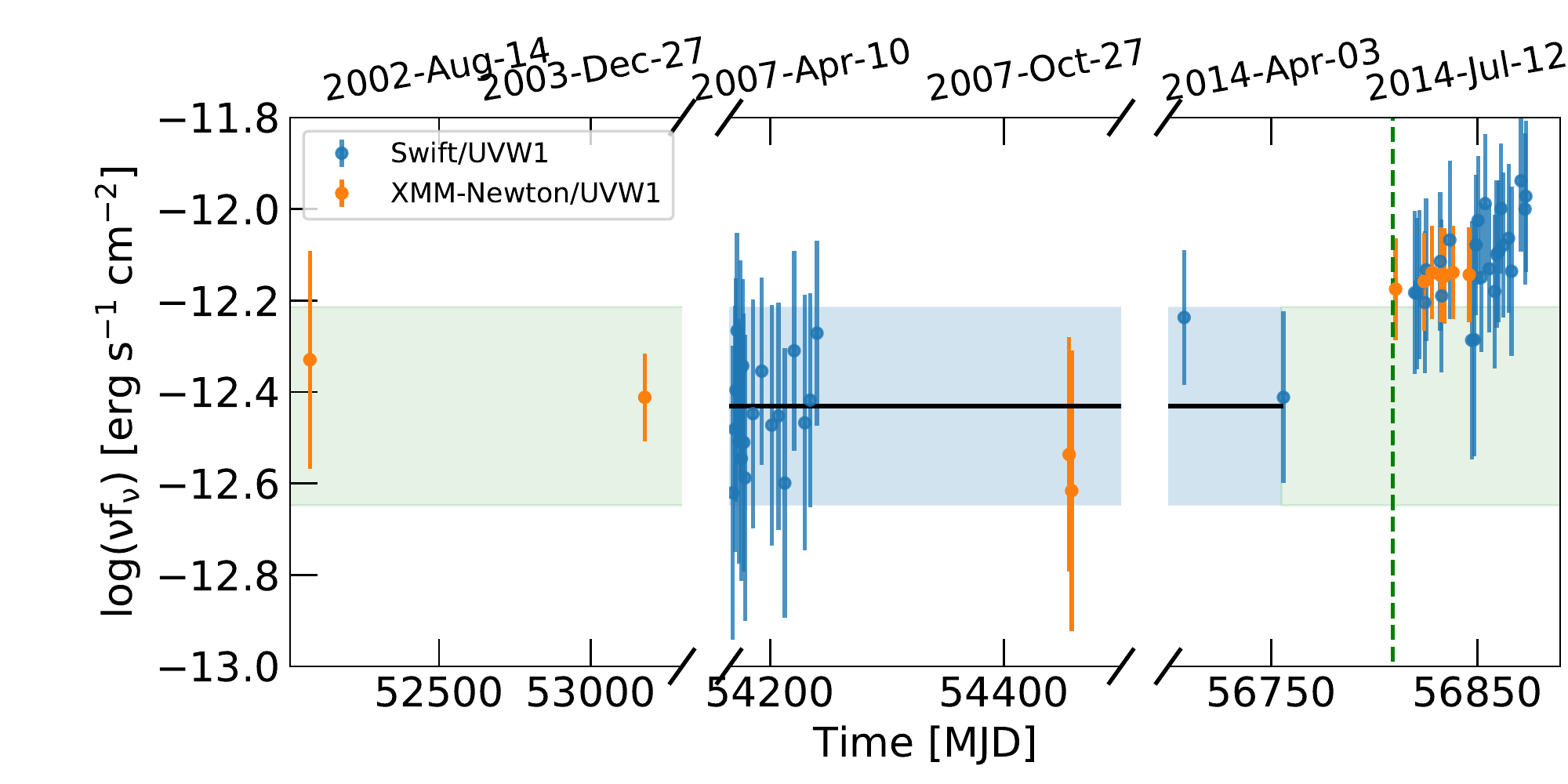}
	\caption{Optical/UV flux light curves of J1214+1402 for Swift and XMM-Newton UVW1 instruments. The black solid line and blue-shaded region correspond to the mean and 2$\sigma$ uncertainty of the photometric flux observed by the Swift/UVW1 instrument in the time range of low optical state. For a better understanding of the variation in the full-time range, the blue-shaded region is extended as green-shaded regions to other time ranges. The green dashed line represents the time of 2014 June 01.}
	\label{fig:xmm_swift}
\end{figure}

\begin{figure}
	\centering
	\includegraphics[width=0.85\columnwidth,angle=0]{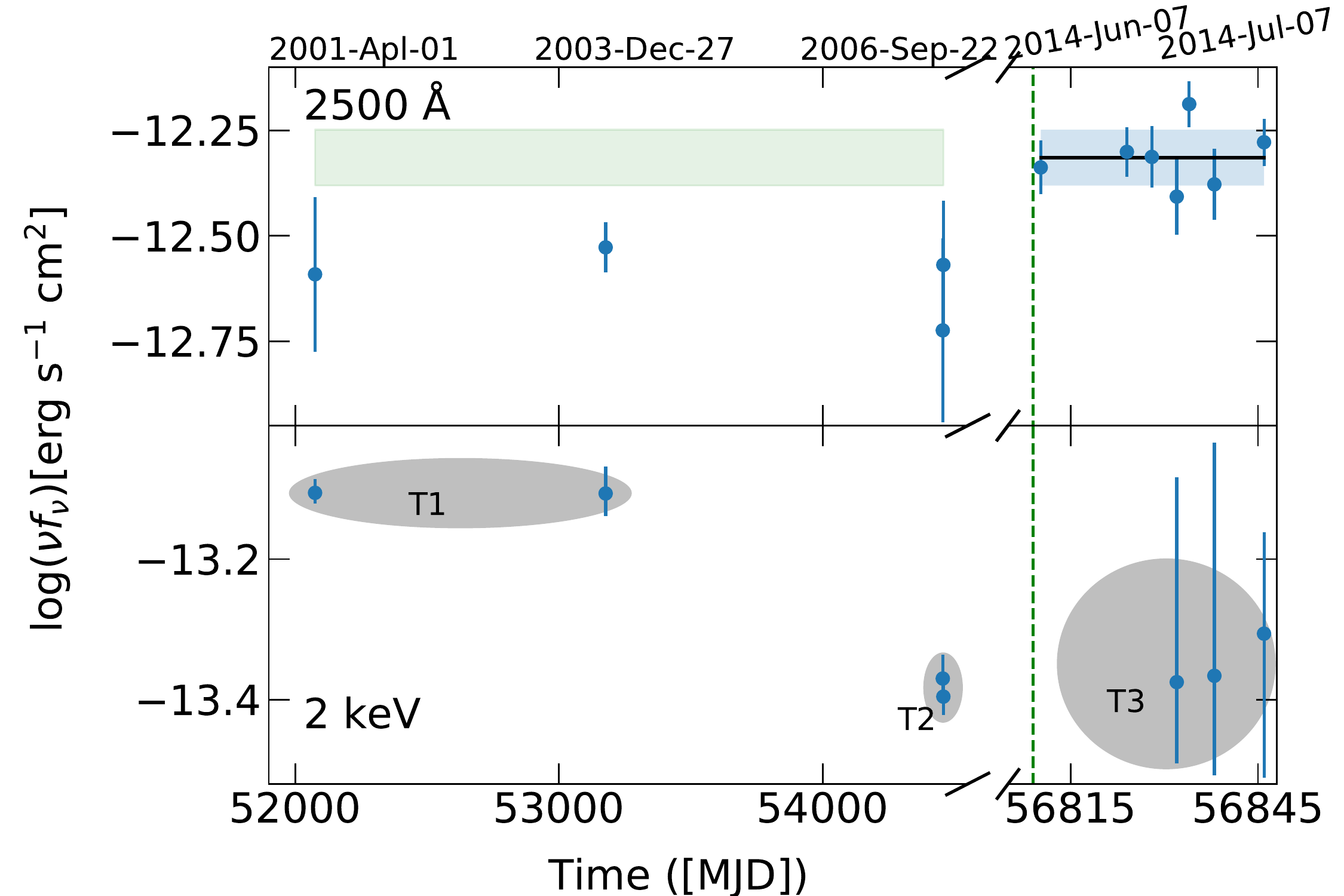}
	\caption{The simultaneous optical/UV and X-ray light curves of J1214+1402 from XMM-Newton observations. The upper and lower panels for the flux at rest frame $2500$ \AA ~and $2\,\rm keV$, respectively. The black solid line and blue-shaded region correspond to the mean and $1\sigma$ uncertainty of the flux at rest-frame $2500$ \AA~ after 2014 June 01. The green dashed line represents the time of 2014 June 01. For a better understanding of the variation in the full-time range, the blue-shaded region is extended as the green-shaded region. The gray ellipses (T1, T2, and T3) encircle the data in three time ranges during XMM-Newton observations.}
	\label{fig:xmm_ltc}
\end{figure}

Figure \ref{fig:swift} shows the optical/UV and X-ray light curves of J1214+1402 based on Swift observations. We compared the optical/UV flux for UVW1 band of Swift and XMM-Newton telescopes in Figure \ref{fig:xmm_swift}. 
Figure \ref{fig:xmm_ltc} shows the optical/UV and X-ray light curves of XMM-Newton data. 
In total, there are 11 available optical/UV photometric observations by XMM-Newton, but only 7 X-ray spectra. Similar to the luminosity $L_{3000}$ calculated above, the flux at rest frame $2500$ \AA~were also calculated from $V$ band count rate by OM instrument, assuming SDSS optical continuum slope $\alpha = -0.94\pm 0.02$. The flux at rest frame $2\rm \ keV$ was obtained by X-ray spectral fitting, and corrected by $0.20~ \rm dex$ if the correlate spectrum was extracted from the circular region of $7.5\arcsec$. 

In Figure \ref{fig:xmm_swift}, consistent with the result of Swift, the UVW1 light curve of XMM-Newton also shows a high optical/UV flux state after 2014 June 1. This is also supported by the light curve of rest frame $2500$ \AA~in Figure \ref{fig:xmm_ltc}. 

\begin{figure}
	\centering
	\includegraphics[width=0.85\columnwidth,angle=0]{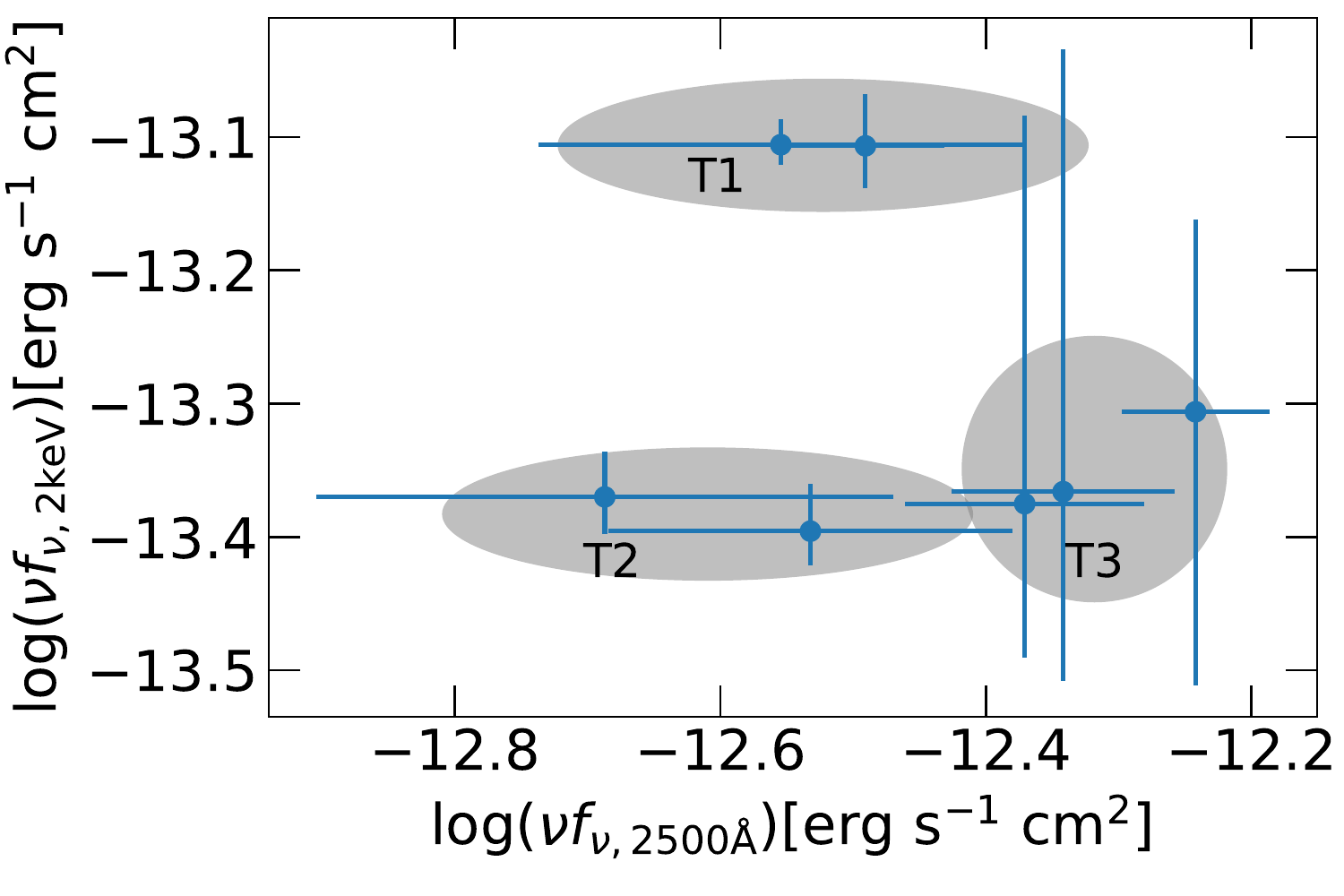}
	\caption{The optical to X-ray flux correlation of J1214+1402 based on simultaneous observations of XMM-Newton. The gray ellipses encircle the data in three time ranges during XMM-Newton observations.}
	\label{fig:xmm_o-x}
\end{figure}

For better understanding, we divide the time duration of XMM-Newton observations into three time ranges, T1 for the first two epochs 2001 June 15 and 2004 June 21, T2 for the epochs of 2007 December 21 and 2007 December 23, and T3 for the epochs after June 2014 (from 2014 June 02 to 2014 July 07). The light curve of X-ray at $2~\rm keV$ in the lower panel in Figure \ref{fig:xmm_ltc} shows no significant flux variation for T2 and T3. This is consistent with the finding from the light curves of Swift in these overlapped time periods. But the X-ray light curve in Figure \ref{fig:xmm_ltc} presents brighter X-ray flux (high X-ray state) in T1 when optical/UV emission stays in low state. In contrast, the X-ray flux stayed at low state for T3, while the optical/UV is at high state.
With simultaneous X-ray and optical/UV observations of XMM-Newton, we plotted the flux relation between $2~\rm keV$ and $2500$ \AA~in Figure \ref{fig:xmm_o-x}. It shows no significant correlation between X-ray and optical/UV bands with all data in three time ranges, with Spearman correlation coefficient $r_{\rm s} = 0.036$. Our results clearly indicate independent optical to X-ray flux variations in our studied period. 

\section{Discussion}
\label{sec:discus}

\subsection{Soft X-ray excess}

The soft X-ray excess emission of RLQs was found to be different from that of RQQs for the lower soft X-ray excess detection \cite[][]{2011MNRAS.417..992S, 2012MNRAS.423.2633S, 2020ApJ...893...39Z} and relatively lower soft X-ray flux in RLQs \cite[][]{2011ApJS..196....2S}. However, \cite{2021RAA....21....4Z} found that RLQs have higher optical normalized composite X-ray spectrum than RQQs at $>99.9\%$ confidence level at both soft and hard X-ray bands, based on a larger sample than that of \cite{2011ApJS..196....2S}. The soft X-ray excess mechanism in AGNs needs more discussion. 

There are several models to explain the soft X-ray excess in AGNs: (1) the model of blurred reflection from ionized accretion disk \cite[][]{2005MNRAS.358..211R, 2006MNRAS.365.1067C}, (2) the thermal Comptonization by a warm optically thick corona surrounding the inner regions of the disk \cite[][]{1998MNRAS.301..179M, 2004MNRAS.349L...7G, 2012MNRAS.420.1848D}, and (3) the possible warm absorber \cite[i.e., partially ionized absorption model, e.g., ][]{1993ApJ...415..129F, 2004MNRAS.349L...7G}. 
The partial covering model can be ruled out in our case for the absence of the absorption edge around $1~\rm keV$ \cite[e.g.][]{1993ApJ...415..129F}, as shown in Figure \ref{fig:xmm_spec}. 

According to the relativistic blurred disk reflection model \cite[][]{2005MNRAS.358..211R}, the accretion disk is illuminated by a power-law X-ray continuum emission and produces a reflected spectrum at about 30 keV, a strong fluorescent Fe K$\alpha$ line at 6.4 keV, and a soft X-ray excess. The relativistically smeared reflection model has been widely used to explain the physical origin of the soft excess in many AGNs, such as 1H 0323+342 \citep[][]{2020MNRAS.496.2922M}; IRAS 09149-6206 \citep[][]{2020MNRAS.499.1480W}; 1H 0707-495 \citep[][]{2021A&A...647A...6B}. For example, \cite{2018MNRAS.473.3584P} suggested that the soft X-ray excess in 1H 0419-557 can be explained by the blurred reflection model with soft excess flux and power-law flux varied by a factor of $\sim 2$ and $\sim 7$, respectively, but the variability amplitudes of optical/UV band were only at $6 \sim 10$ percent level.
In the case of J1214+1402 with XMM-Newton observations, as in Table \ref{tab:xmm_xfit}, we find the soft X-ray excess below $1~\rm keV$ only appears at high optical state (T3). The hard X-ray slope roughly remains constant ($\Gamma \simeq 1.8$) for all spectra, and so, the light curve of $2~\rm keV$ flux can represent the variation of hard X-ray emission. 
Thus, the light curve in Figure \ref{fig:xmm_ltc} suggests that J1214+1402 has no prominent hard X-ray flux variation between T2 and T3, when optical flux changes, but has variable soft X-ray excess. 
Moreover, the two spectra in T1 with the highest hard X-ray flux have no significant soft X-ray excess. 
We conclude the soft X-ray excess emerges independently of the hard X-ray emission. 
In the reflection model, the soft X-ray emission is expected to be less variable than the hard X-ray component \cite[][]{2011A&A...534A..39M} and should be correlated with the hard X-ray emission \citep[e.g., ][]{2007ApJ...671.1284D, 2013MNRAS.436.3173J}. It is therefore hard to explain the origin of the soft X-ray excess in J1214+1402 with the relativistic reflection model. 
The X-ray spectra do not show any reflection features, such as the Fe K$ \alpha $ line (see Figure \ref{fig:xmm_spec}). Although the intensity of Fe K$ \alpha $ line may depend on the disk inclination \cite[][]{2005MNRAS.358..211R}, strong reflection component and iron lines have been found in AGNs with disk inclination between $20\degr$ to $60\degr$ \citep[e.g.,][]{2005astro.ph..7409F, 2005MNRAS.361.1197C, 2006AN....327.1079R, 2006A&A...445L...5P}, the inclination range applicable to our source as a steep-spectrum radio quasar. Thus the result of no visible reflection feature is further against the reflection model in J1214+1402.

In RQ-AGNs, it is believed that the primary X-ray continuum originates from optically thin ($\tau \sim 1$) high temperature ($\sim 100~\rm keV$) ``corona'' \cite[e.g.,][]{2015A&ARv..23....1B}. Apart from the corona, the warm corona model assumes that an optically thick ($\tau \sim 10-20$), low temperature ($\sim 0.1-1.0~\rm keV$) plasma inverse Compton scatters the optical/UV photons to the soft X-ray band \cite[e.g.,][]{2012MNRAS.420.1848D}. The X-ray spectra of many AGNs were proved to be well described by the warm corona model, such as Zw 229.015 \citep[][]{2019MNRAS.488.4831T}, Ton S180 \citep[][]{2020MNRAS.497.2352M}, and SBS 1353+564 \cite[][]{2021MNRAS.507.3572X}.

Many works on RQQs \cite[e.g.,][for NGC 5548, and Mrk 509, respectively]{1998MNRAS.301..179M, 2011A&A...534A..39M} found that the soft X-ray excess varies in association with the optical/UV emission and the origin of soft excess can be explained as the result of warm Comptonisation of the disk emission. \cite{2011A&A...534A..39M} found that the change in power-law X-ray component of Mrk 509 is smaller than that of the soft X-ray excess and is not correlated with the flux variability of soft X-ray excess and disk component on the probed timescale. \cite{2013MNRAS.436.3173J} showed that the fast variability of the soft X-ray excess in PG 1244+026 is independent of the $4-10~\rm keV$ variability. 
Similar to these studies, the XMM-Newton data of J1214+1402 shows high X-ray flux but low optical flux at the beginning, and the X-ray flux changes to and remains at a low state afterward irrespective of optical variations. The soft X-ray excess appearing only during high optical state suggests the presence of a warm corona in J1214+1402. As the optical/UV luminosity rise, the soft X-ray emission increases and exceeds the hard power-law spectrum. Otherwise, the soft X-ray excess decreases and can be easily dominated by the primary power-law component. 

Our results seem to support the warm corona to be the origin of the soft X-ray excess in RLQ J1214+1402. 

\subsection{Hard X-ray emission}

In the disk-corona model, the primary power-law X-ray is believed to be produced from a hot corona near to the supermassive black hole by the inverse Compton scattering of optical/UV photons \cite[e.g.,][]{1980A&A....86..121S, 1993ARA&A..31..717M, 2015A&ARv..23....1B}. 
In AGN multi-band variability studies, the optical-to-X-ray time lag was found to be about intra-day to few days \cite[e.g.,][]{2009MNRAS.397.2004A, 2014MNRAS.444.1469M, 2018MNRAS.480.2881M}. If the short time-scale optical/UV variability is driven by reprocessing of X-rays by a surrounding accretion disk, it would be reasonable to argue that the corona locates close to the disk with a distance of about a few light days. Moreover, the measurements based on both X-ray reverberation mapping and gravitational microlensing suggest that the corona is quite compact (less than 20 $R_{\rm g}$, were $R_{\rm g} \equiv GM_{\rm BH}/c^2$ is the gravitational radius) around the supermassive black hole \cite[][]{2012ApJ...756...52M, 2013ApJ...769...53M, 2013ApJ...769L...7R, 2014MNRAS.438.2980C}. 

Combining the Swift and XMM-Newton results, the optical/UV light curve of J1214+1402 clearly presents two states, a high optical/UV flux after 2014 June 1 and a low state before 2014 April 8, but the hard X-ray emission shows no significant variation after 2007 March (including T2 and T3, see Figures \ref{fig:swift} and \ref{fig:xmm_ltc}). Moreover, in 2001 June and 2004 June when J1214+1402 stayed in low optical state, the X-ray light curve showed the brightest emission. 
All X-ray spectra extracted from XMM-Newton data of J1214+1402 have similar X-ray photon index considering the error bars. And so there is no significant hard X-ray slope change with the increases in the Eddington ratio.
If the hard X-ray emission of J1214+1402 arises from the disk-corona system, due to reprocessing \citep[][]{1988MNRAS.233..475G} or magnetic turbulence \citep[][]{2018ApJ...868...58K}, the primary power-law X-rays should vary together with optical/UV emission in a few days considering the distance between the corona and the disk being a few light days \citep[e.g.,][]{2018ApJ...860...29Z} or $<20~R_{\rm g}$ \citep[about 3.87 light days in our case, e.g.,][]{2013ApJ...769L...7R}.
These results seem hard to be explained by the disk-corona model \cite[see also, e.g., ][]{2018MNRAS.474.5351P} due to the independent optical-to-X-ray variations in the probed time range in our work.  

Extensive works have been carried out to study the multi-wavelength flux variability in RQQs, from which the X-ray emission is believed to be originated from the disk-corona system \cite[e.g.,][and references therein]{2015A&ARv..23....1B}. In contrast, to our best knowledge, there are only a few multi-band variability studies on non-blazar RLQs.
\cite{2018ApJ...866..132B} presented the long-term multi-wavelength monitoring of the broad line radio quasar 4C +74.26. Employing discrete cross-correlation function (DCF) analysis, \cite{2018ApJ...866..132B} found that the optical emission lags behind the radio and X-rays by $250 \pm 42$ and $105\pm 58$ days with the global significance of 98\% and 87\%, respectively. The authors discussed the disk-jet connection based on optical-radio variation but paid less attention to the X-ray origin. We re-analyzed the hard X-ray count rates variation of Figure 1 in \cite{2018ApJ...866..132B}, and found the excess variance $\sigma_{\rm rms}^2 = 1.87\rm E -08$ with mean value $0.0005\pm 0.0002$. It indicates no significant X-ray flux variation when the optical flux changes, consistent with the low significance level (<90\%) of the time lag between optical and X-rays. This result of no strong correlation between optical and X-ray flux of 4C$+$74.26 is similar to that of our source J1214+1402. 

It is widely believed that the optical emission in blazars is usually dominated by non-thermal process from relativistic jet \cite[e.g.,][and references therein]{1995PASP..107..803U, 1997ARA&A..35..445U, 2012MNRAS.425.3002G, 2020Galax...8...58K}. In some cases, the thermal emission is significant, for example, the significant BBB in the optical/UV band was found in the well-known blazar, 3C 273 \citep[e.g.,][]{2005ApJ...619...41S}.
\cite{2015MNRAS.451.1356K} studied the optical/UV and X-ray variability of 3C 273, and found no relationship between the X-ray and optical/UV emission. The authors support the scenario where two independent particle populations responsible for the optical/UV and X-ray emission respectively are present in the local emitting regions. 
Similar to our source, the optical/UV emission of 3C 273 also shows a prominent BBB and varies independently with the X-rays. The presence of BBB indicates that 3C 273 and J1214+1402 may have similar optical/UV origination which could be dominated by the thermal emission of the accretion disk. In contrast, the X-ray emission of 3C 273 is very complicated, such as the two-components scenario with a Seyfert-like and a blazar-like component \cite[][]{2004Sci...306..998G} or inverse Compton processes (SSC and/or EC) from the base of the jet \cite[][]{2008A&A...486..411S}, but more or less be associated with the relativistic jet. 
Considering the similarity of independent X-ray-to-UV/optical variability and thermal UV/optical emission between 3C 273 and our target, the X-ray emission of J1214+1402 can possibly be related with jet.

Many sample studies presented that the jet may also play an important role in the X-ray production for non-blazar AGNs, such as, \cite{2011ApJ...726...20M, 2015ApJS..220....5M, 2017ApJ...835..226K, 2020MNRAS.497..482L, 2020ApJ...893...39Z, 2021RAA....21....4Z}. 
\cite{2017ApJ...835..226K} found an inverse correlation between reflection fraction and radio Eddington luminosity, and a positive correlation of the path length connecting the corona and reflecting regions of the disk to the radio Eddington luminosity. 
These correlations can be well explained by their corona-jet model. 
\cite{2018ApJ...866..132B} explored the disk-jet connection in the broad-line radio quasar 4C +74.26 and found a good correlation between the optical and radio bands with the disk lagging behind the jet by $250\pm 42$ days. The authors argued the lag may be related to a delayed radiative response of the disk when compared with the propagation timescale of magnetic perturbations along a relativistic outflow.
For the lack of simultaneous long-term optical-to-radio data, it is hard to estimate the variation time lag between optical and radio bands in J1214+1402. However, J1214+1402 shows similar properties with 4C +74.26, e.g., broad emission lines in the optical spectrum, strong radio emission, and no distinct correlation in the variation between optical/UV and X-ray bands. 
Analogous to 4C +74.26, if the X-ray emission of J1414+1402 is related to the jet, the optical-to-radio time delay in \citet[][more than 200 days time delay in 4C +74.26]{2018ApJ...866..132B} may give a possible explanation for our optical and X-ray light curves.

Changing-look AGNs (CL-AGN) are a special class of AGNs with broad emission lines observed to appear or disappear together with large changes in continuum luminosity. Recently, more and more works suggest that the variation of the accretion rate is likely to be the primary origin for CL-AGNs \citep[e.g.,][]{2018ApJ...866..123M, 2019ApJ...883L..44G, 2020ApJ...898L...1R}. And the intrinsic mechanism for their variable accretion could be different in each case, like accretion state transition \citep[e.g.,][]{2018ApJ...866..123M, 2019ApJ...883...76R} or tidal disruption event \citep[TDE, e.g.,][]{2015MNRAS.452...69M, 2020ApJ...898L...1R}. There is some CL-AGNs show similar optical to X-ray variation with J1214+1402 in some respects. 
Mrk 590 is a well-known CL-AGN, and its soft X-ray excess vanished in 2011 January \citep[][]{2012ApJ...759...63R} and reappears with the increase of optical/UV flux in 2014 November \citep[][]{2018ApJ...866..123M, 2019MNRAS.486..123R}. However, different from J1214+1402, the hard X-ray power-law component of Mrk 590 varies together with the optical/UV emission \citep[][]{2014ApJ...796..134D}. \cite{2018ApJ...866..123M} suggested Mrk 590 may have been in a transition state for the dramatic variation in the accretion rate. 
The Eddington ratio of J1214+1402 varies from 0.033 to 0.114 (see Table \ref{tab:xmm_omcts}), all above the transition value of 0.01 between hot accretion flow and cold accretion flow \citep[e.g.,][]{2014ARA&A..52..529Y}. Considering all these facts, the changing-look scenario (i.e. accretion transition) in J1214+1402 is less likely. 

CL-AGN 1ES 1927+654 has significant optical flux increases on 2017 December 23. After the optical/UV outburst ($t \sim 160$ days), \cite{2020ApJ...898L...1R, 2021ApJS..255....7R} presented the long-term X-ray monitoring on the source, and found the first X-ray observation carried in 2018 May has a similar X-ray luminosity level of the previous observation in 2011 May, and found the X-ray power-law emission significantly decreased or even disappeared in 1ES 1927+654 in the following $\sim 40$ days and then increased in luminosity but varies independently with optical/UV flux. The authors speculated that the characteristics of 1ES 1927+654 might have resulted from the interaction between a tidally disrupted star with an accretion disk around a central black hole, which would empty the innermost regions of the accretion flow. 
Similar to 1ES 1927+654, J1214+1402 has a significant increase in the optical/UV band flux, while the X-ray remains relatively stable at Swift observations. 
If the flux increases in the optical/UV band in J1214+1402 originates from the TDE event, the X-ray may vary after the optical/UV burst $>100$ days \citep[see also][]{2021ApJ...912..151W}. However, the possibility of TDE event is excluded due to the heavy black hole mass in J1214+1402 ($M_{\rm BH} \approx 10^{9.23} \rm M_\odot$) \citep[e.g.,][]{2020SSRv..216...85S, 2021A&A...652A..15H}.

It should be noticed that all our analysis are based on sparse data in limited time coverage, therefore, the long-term multi-wavelength monitoring of J1214+1402 with good sampling will be needed in the future to further study the physical mechanism of multi-band emission and the related variabilities.

\section{Summary}
\label{sec:sum}

In this work, to explore the X-ray emission mechanism of AGNs, we studied the optical/UV to X-ray flux relation of an RLQ J1214+1402 using Swift and XMM-Newton observations. 
We found that RLQ J1214+1402 should not be a blazar based on the historical multi-band data. 
The Swift data showed that J1214+1402 has two prominent optical/UV states with high flux after 2014 June, but has no significant X-ray variation between these two states, as shown in Figure \ref{fig:swift}. 
These findings were also supported by the XMM-Newton optical/UV and X-ray light curves in the overlapped time period with Swift. 
Moreover, before the time of Swift observations, the XMM-Newton light curve displayed two unusual data in which the source stayed at low optical state but with bright X-ray emission. The independent optical-to-X-ray variations on the probed timescale are hard to be described by the disk-corona system.
The soft X-ray excess appears only at high optical state when the source has no significant variation at hard X-ray. 
Our results suggest that the soft X-ray excess in J1214+1402 should not be related to the blurred reflection model and may be explained with the warm corona model.

\section*{Acknowledgements}

We thank the anonymous referee for valuable and insightful suggestions that improved the manuscript.
This work is supported by the Science and Technology Project funded by the Education Department of Jiangxi Province in China (Grant No. GJJ211733), the Doctoral Scientific Research Foundation of Shangrao Normal University (Grant No. K6000449), the National Science Foundation of China (grant 11873073), Shanghai Pilot Program for Basic Research $-$ Chinese Academy of Science, Shanghai Branch (JCYJ-SHFY-2021-013), and the science research grants from the China Manned Space Project with NO. CMSCSST-2021-A06.

Part of this work is based on archival data, software or online services provided by the Space Science Data Center - ASI.
This research has made use of the NASA/IPAC Extragalactic Database (NED), which is funded by the National Aeronautics and Space Administration and operated by the California Institute of Technology. 
This research has made use of data and/or software provided by the High Energy Astrophysics Science Archive Research Center (HEASARC), which is a service of the Astrophysics Science Division at NASA/GSFC.
This work is based on results from the enhanced XMM-Newton spectral-fit database, an ESA PRODEX funded project, based in turn on observations obtained with XMM-Newton, an ESA science mission with instruments and contributions directly funded by ESA Member States and NASA. 
Funding for SDSS-III has been provided by the Alfred P. Sloan Foundation, the Participating Institutions, the National Science Foundation, and the U.S. Department of Energy Office of Science. The SDSS-III web site is http://www.sdss3.org/. 

\section*{Data Availability}

The data underlying this article are available in NRAO (https://data.nrao.edu/portal), XMM-Newton (https://www.cosmos.esa.int/web/xmm-newton) and Swift (https://www.swift.ac.uk/index.php) data centers.  



\bibliographystyle{mnras}
\bibliography{ms} 




%
%


\bsp	
\label{lastpage}
\end{document}